%% file: preprint.tex
\documentstyle[emulateapj,epsf]{article}
\received{}
\accepted{}
\journalid{}{}
\articleid{}{}
\newcommand {\ks} {km~s$^{-1} \;$}
\newcommand {\h} {$\ h^{-1} \, Mpc \;$}
\begin{document}
\title{The  Observational  Distribution of Internal  Velocity Dispersions\\
in Nearby Galaxy Clusters}
\author{
D. Fadda$^{1}$, M. Girardi$^{1}$ \\
G. Giuricin$^{1}$, F. Mardirossian$^{1,2}$, and M. Mezzetti$^{1}$}
\affil{
$^1$Dipartimento di Astronomia,
Universit\`{a} degli Studi di Trieste, \\
SISSA, via Beirut 4, 34013 - Trieste, Italy, \\
$^2$Osservatorio di Trieste, via Tiepolo 11, 34100 - Trieste, Italy; \\
Email:  fadda, girardi, giuricin, mardirossian, mezzetti @sissa.it}

\begin{abstract}
 
We analyze the internal velocity dispersion, $\sigma$, of a sample of
172 nearby galaxy clusters (z$\le0.15$), each of which has at least
30 available galaxy redshifts, and spans a large richness
range. Cluster membership selection is based on nonparametric
methods. In the estimate of galaxy velocity dispersion we consider
the effects of possible velocity anisotropies in galaxy orbits, the
infall of late-type galaxies, and velocity gradients. The dynamical
uncertainties due to the presence of substructures are also taken into
account.

Previous $\sigma$-distributions, based on smaller cluster samples, are
complete for the Abell richness class $R\ge 1$.  In order to improve
$\sigma$ completeness, we enlarge our sample by including also
poorer clusters.  By resampling 153 Abell-ACO clusters, according to
the richness class frequencies of the Edinburgh-Durham Cluster
Catalog, we obtain a cluster sample which can be taken as
representative of the nearby Universe.

Our cumulative $\sigma$-distribution agrees with previous
distributions within their $\sigma$ completeness limit ($\sigma
\gtrsim$ 800 \ks).  We estimate that our distribution is complete for
at least $\sigma \ge 650$ \ks. In this completeness range, a fit of
the form $dN\propto \sigma^{\alpha}d\sigma$ gives
$\alpha=-(7.4^{+0.7}_{-0.8})$, in fair agreement with results coming
from the X-ray temperature distributions of nearby clusters.

We briefly discuss our results with respect to $\sigma$-distributions
for galaxy groups and to theories of large scale structure formation.

\end{abstract}

\section{Introduction}

Galaxy clusters, which are the most massive bound galaxy systems and
which have collapsed very recently or are just collapsing, play an
important role in the study of large scale structure formation. In
particular, the distribution of internal velocity dispersion in galaxy
clusters can be a strong constraint of cosmological models (e.g. Frenk
et al. 1990; Bartlett \& Silk 1993; Jing \& Fang 1994; Crone \& Geller
1995).

The velocity dispersion (hereafter $\sigma$) of a galaxy population,
which is in dynamical equilibrium within the cluster and traces the
whole system, is directly linked to the total gravitational potential
via the virial theorem. The precise relation between mass and
dispersion depends on an assumption about the relative distribution of
mass and galaxies (Merritt 1987).

The observational determination of the $\sigma$ distribution
encounters several problems.  Some of these arise in the estimate of
$\sigma$ and are due to, e.g.,cluster member selection, velocity
anisotropy in galaxy orbits, cluster asphericity, possible infall of
spiral galaxies into the cluster, and the presence of substructures
(see e.g. Girardi et al. 1996, hereafter G96, for a detailed
discussion of these topics).

In particular, cluster velocity anisotropies are poorly known
(e.g. Merritt 1987; Dejonghe 1987). In order to avoid effects of
possible anisotropies on $\sigma$ estimates, G96 suggested studying
the "integral" velocity dispersion profile (hereafter VDP), where the
dispersion at a given radius is evaluated by using all the galaxies
within that radius.  Although the presence of velocity anisotropies can
strongly influence the value of $\sigma$ computed for the central
cluster region, it does not affect the value of the spatial (or
projected) $\sigma$ computed for the whole cluster (The \& White 1986;
Merritt 1988). Observationally, the VDPs of several clusters show
strongly increasing or decreasing behaviours in the central cluster
regions, but they are flattening out in the external regions (beyond $\sim
$1 \h) suggesting that in such regions they are no longer affected by velocity
anisotropies (Figure~1 of G96).  Thus, while the $\sigma$-values
computed for the central cluster region could be a very poor estimate
of the depth of cluster potential wells, we can reasonably adopt the
$\sigma$ value computed by taking all the galaxies within the radius at
which the VDP becomes roughly constant.

As a general result of their analysis, G96 found evidence of a fair
equipartition between galaxy and gas energy, which suggests that
$\sigma$, like the X-ray temperature of the intracluster medium, can be
a good estimate of cluster potential wells.

Other problems regarding the determination of $\sigma$-distribution arise
from the need to have a cluster sample representative of the
Universe, at least above a lower limit of $\sigma$.

The present $\sigma$-distributions are based on cluster samples which are
complete with respect to cluster richness but not with respect to
$\sigma$ (Frenk et al. 1990; Girardi et al. 1993; Zabludoff et
al. 1993a [Z93]; Collins et al. 1995; Mazure et al. 1996 [M96]).  

The samples used are generally complete for the Abell richness class $R\ge
1$ or for $R$ intermediate between 0 and 1 (Abell counts $N_c \sim
40$; Collins et al. 1995).  Since the correlation between $N_c$ and
$\sigma$ is very broad (e.g. Girardi et al. 1993, M96), the
completeness of a sample with respect to a certain value of richness does not
imply completeness of $\sigma$. M96 estimated that their cluster
sample is complete for $\sigma \ge 800$ \ks. However, they did not
have any $R<1$ cluster in their sample. So the bias introduced by the
absence of poor clusters ($R\le0$), which is common to all
previous works, calls for further analysis.

Moreover, the cluster number density is not well known and the
estimate of it varies from author to author (see M96 and references
therein).

The aim of this work is to obtain a $\sigma$-distribution based on a
large cluster sample which considers also poor clusters, each cluster
having a reliable $\sigma$ estimate.  In order to obtain good
estimates of $\sigma$, we have adopt the procedure of G96, introducing some
improvements described in detail in \S~2. Hereafter we indicate by
$\sigma$ the line-of-sight velocity dispersion.

In \S~2 we describe the data-sample and our selection procedure for
cluster membership; in \S~3 we compute our values of velocity
dispersions; clusters with substructures are analyzed in \S~4; in \S~5
we obtain our $\sigma$-distribution; in \S~6 we discuss our results
and draw our conclusions.

All the errors are given at the 68\% confidence level (hereafter c.l.).

A Hubble constant of 100 $h^{-1}$ \ks $Mpc^{-1}$ is used.

\section{The Data Sample}

We considered 172 nearby clusters (z$\le$0.15), each cluster having at
least 30 galaxies with available redshift and showing a significant
peak (see \S~2.1) in the redshift space. Moreover, we considered only
galaxy clusters having an error on $\sigma~\lesssim 150$~\ks (for
$\sigma$ computed using all the galaxies within 1\h).  Actually, the
VDPs of clusters showing larger $\sigma$ errors are too noisy for us to
understand their behaviour (\S~3).

\includegraphics{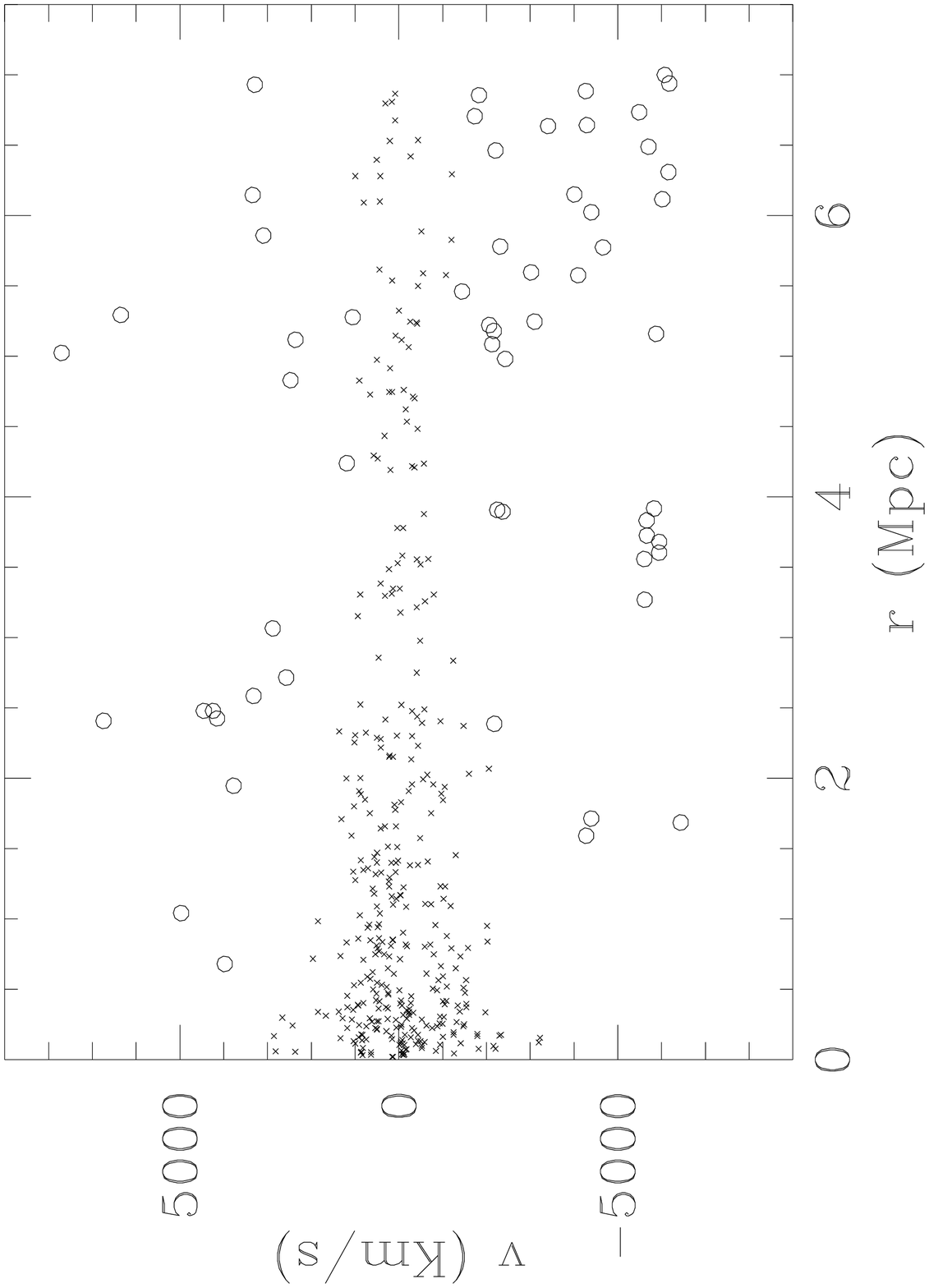}
\vspace{7.2cm}
\vspace{1mm}
{\small\parindent=-3.5mm
{\rm Fig.}~1.---The shifting gapper method applied to the Coma
cluster (A1656).  The velocities are referred to the mean cluster
velocity.  The empty circles indicate the galaxies rejected by our
procedure.}
\vspace{5mm}

We combined data coming from the literature with new data of the ESO
Nearby Abell Clusters Survey (ENACS, Katgert et al. 1996), kindly
provided by the ENACS team. Most of the selected clusters (155) are
Abell-ACO clusters (Abell, Corwin, \& Olowin 1989).  In order to
achieve a sufficiently homogeneous sample, the galaxy redshifts in
each cluster have usually been taken from one reference source or, if
different sources were used, only when the data-sets proved to be
compatible.  Table~1 lists all the 172 clusters considered. In
Col.~(1) we list the cluster names; in Col.~(2) the number of galaxies
with measured redshift in each cluster field; in Col.~(3) the Abell
richness class for each cluster, respectively; and in Col.~(4) the
redshift references.  The Supplementary Clusters in ACO catalog with
$N_c<30$ are classified as belonging to $R=-1$.

Throughout our work we applied homogeneous procedures to the analysis
of the redshifts of the clusters selected. We used robust mean and
scale estimates (computed via the ROSTAT routines -- see Beers, Flynn,
\& Gebhardt 1990), applying the relativistic correction and the usual
correction for velocity errors (Danese, De Zotti \& di Tullio 1980).
When the correction for velocity errors leads to a negative value of
$\sigma$, we adopted $\sigma=0$ with an error equal to the value of
the correction.  Moreover, in several analyses we used the adaptive
kernel technique in one- and two-dimensions, which is a non-parametric
method for detecting and analyzing galaxy systems (see Pisani 1993,
1996 and Appendix~A in G96 for details). Here we may point out that,
in the analysis of the velocity distribution, the method can give the
significance and the position of each detected peak, as well as an
estimate of overlapping between two contiguous peaks.  For the sake of
homogeneity, we preferred to use optical centers, which can be
computed by using the two-dimensional kernel method for all clusters,
at every stage of the cluster-membership selection procedure.  The
maps given by G96 (Figure~1) show good agreement between these
optical centers and the X-ray centers. Moreover, the $\sigma$ computed at
large radii are not strongly affected by different choices of cluster
center.

\subsection{Selection Procedure for Cluster Membership}

In determining cluster membership, we first used position and velocity
information sequentially; then we used the two sets of data combined.

The data samples of a few clusters (A2634, A2666, A3556, A3558, and
A4038) encompass large regions of the sky. So we extracted these clusters
by selecting all galaxies within a fixed radius from the cluster center
(2 \h for clusters A2634 and A3558, and 1 \h for clusters A2666,
A3556, A4038). In the case of clusters A399-A401 and A3391-A3395,
which appear very close to one another, we selected galaxy clusters
by considering their respective peaks (obtained via a two-dimensional
adaptive kernel analysis). Clusters A2063 and MKW3S appear
separated both in space and in redshift, according to the adaptive
kernel method. We chose to separate them in redshift.

Dressler \& Schectman (1988b) found that cluster A3716 is divided into
two clumps (N and S clumps).  Since these clumps, also evidenced by
our analysis, are more distant than 1.5 \h, we decided to separate
them. Cluster A548 shows two clumps (NE and SW clumps in Dressler~\&
\pagebreak
$\!$
\includegraphics{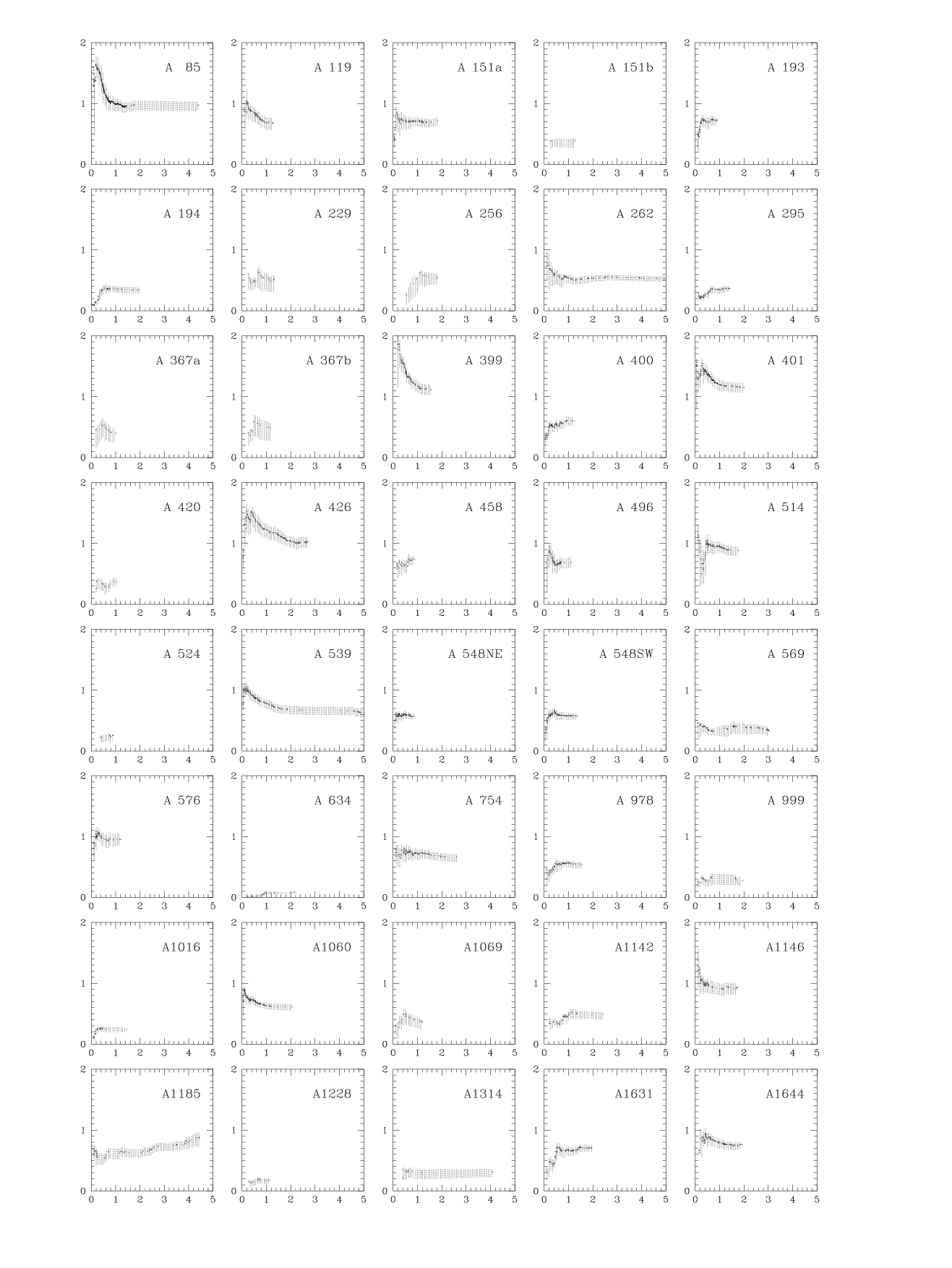}
\vspace{24cm}
\newpage
$\!$
\includegraphics{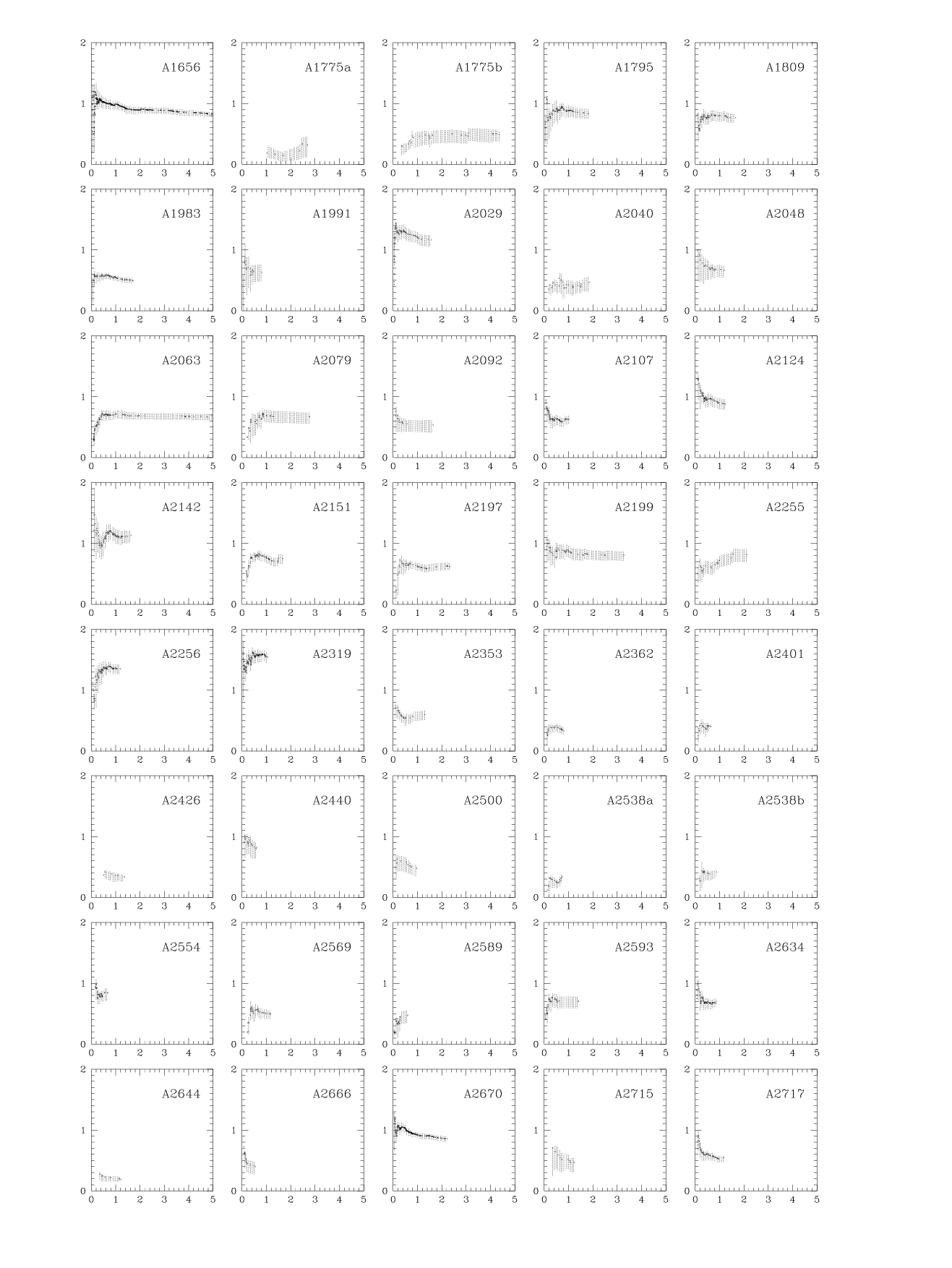}
\vspace{24cm}
\newpage
$\!$
\includegraphics{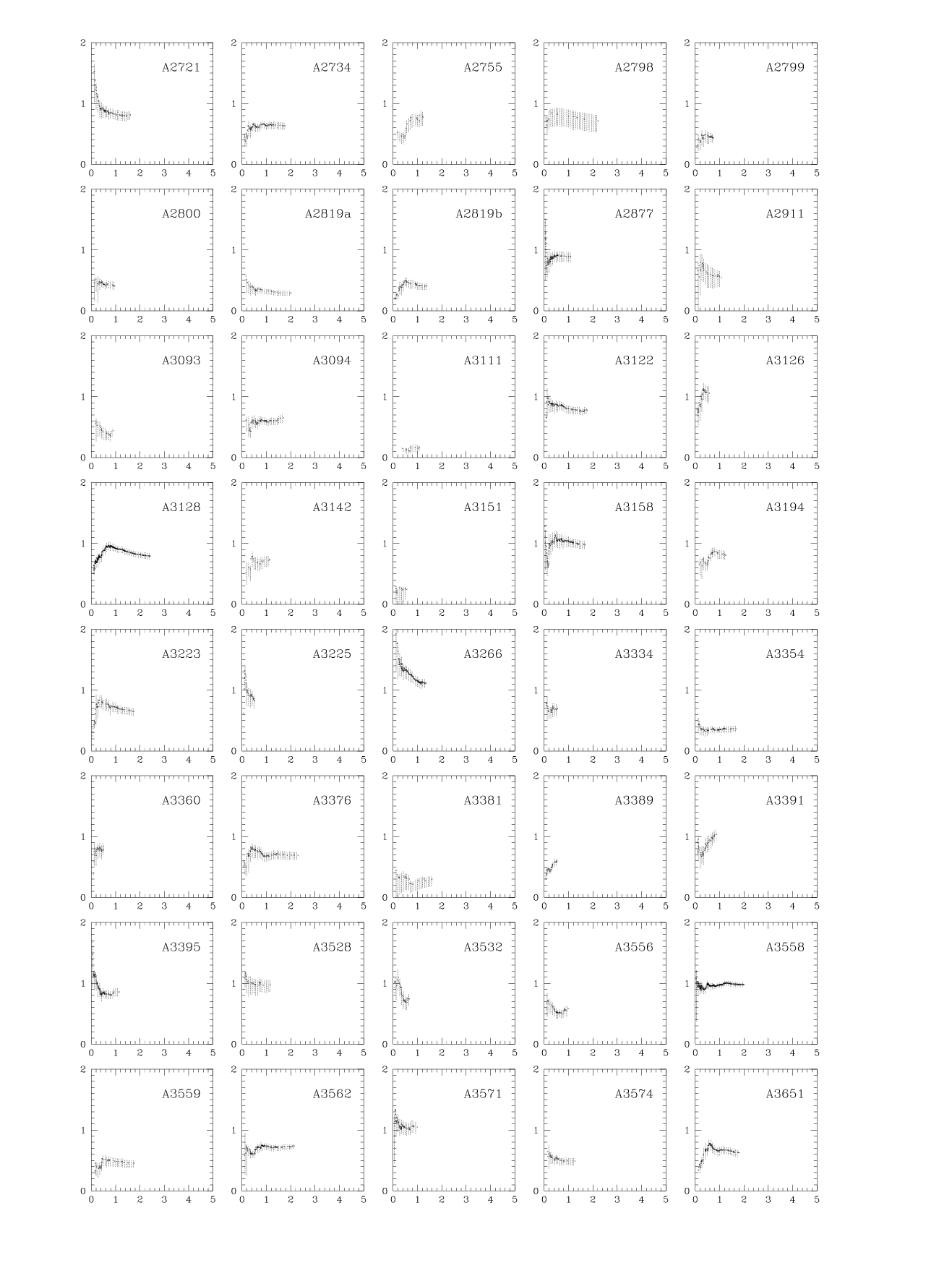}
\vspace{24cm}
\newpage
$\!$
\includegraphics{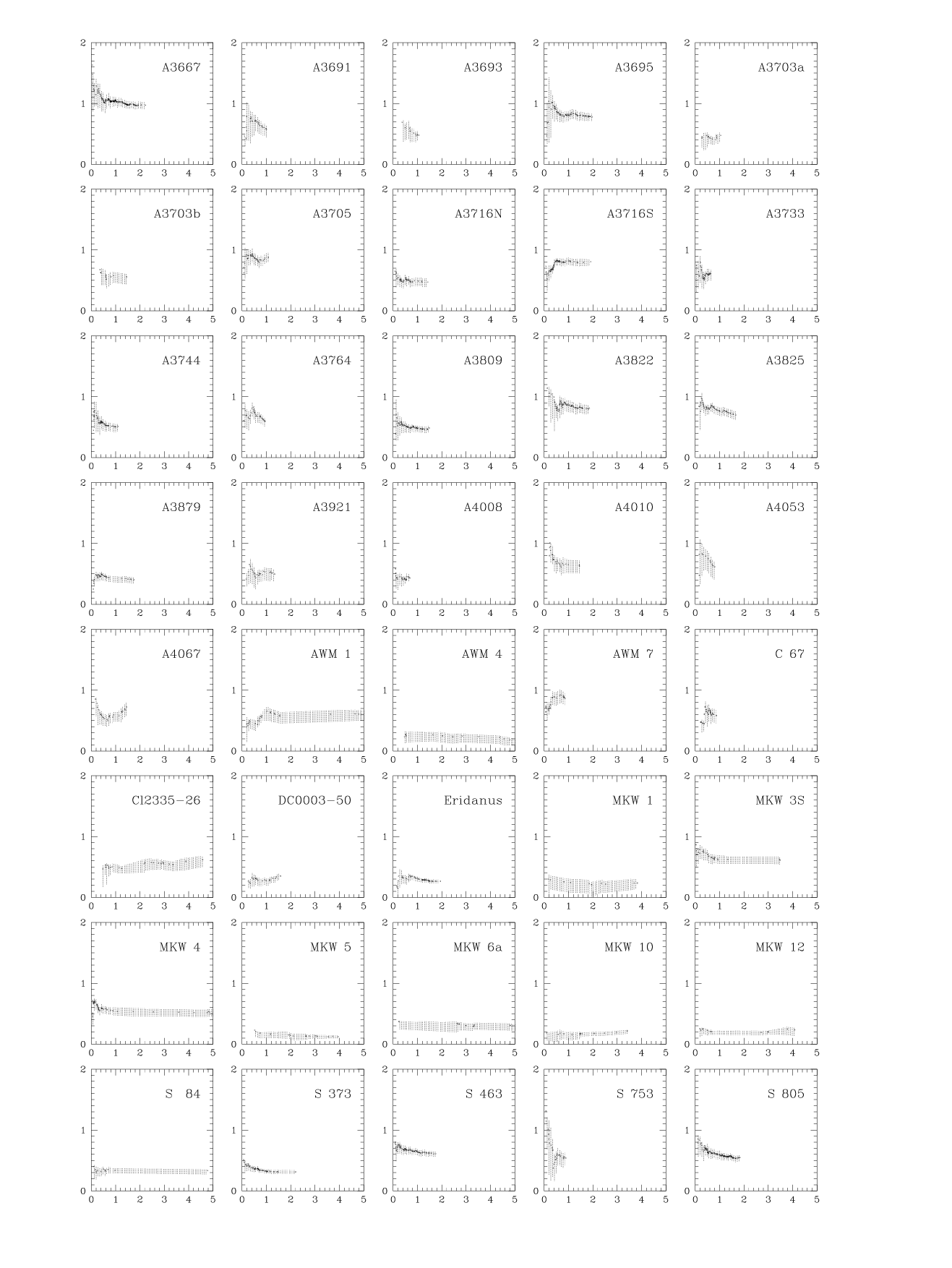}
\vspace{24cm}
\newpage
\includegraphics{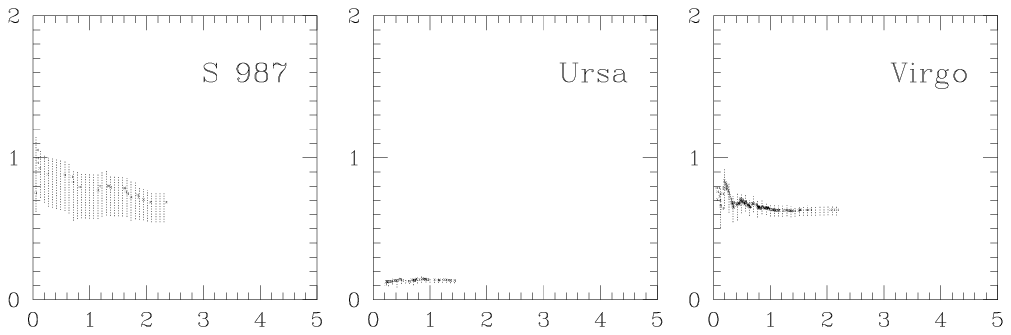}
\begin{minipage}{8.5cm}
\vspace{3.5cm}
\small\parindent=-3.5mm
{\sc Fig.}~2.---The velocity dispersion profiles (VDP), where
the dispersion at a given radius is the average l.o.s. velocity
dispersion within this radius.  The bootstrap error bands at the 68\%
c.l. are shown.  The distances on the x-axis are in Mpc and the
velocity dispersion on the y-axis are in $10^3$ \ks.
\end{minipage}
\vspace{5mm}
Schectman 1988b) separated by more than 1.5 \h; this fact is
confirmed by X-ray data (Davis et al. 1995). Since the cluster also shows
two peaks in the redshift distribution, we prefer to separate the
two systems in redshift.  The inspection of two-dimensional maps,
produced via the adaptive kernel technique, does not show other obvious
cases of clusters which should be subdivided.

There are several methods available in the literature for performing
the usual cluster-membership selection in velocity space.  For
instance, the fixed gap method rejects all the galaxies separated by
more than a fixed value (e.g. 1000 \ks, as used by Katgert et
al. 1996) from the central body of the velocity distribution. The
weighted gap method uses gaps weighted by distance from the
central body (e.g. Beers et al. 1990; Girardi et al. 1993). The
adaptive kernel method (Pisani 1993) is more sensitive to the presence
of peaks in the velocity distribution. For instance, the cluster
fields of A3526 and A539 show two peaks according to the adaptive
kernel method (Pisani 1993), but the peaks in the field of A3526 are not
detected by the two previous methods and the fixed gap method fails also in
the case of A539.

We decided to adopt the adaptive kernel method, considering only
clusters which show a significant peak (at least at the 99\%
c.l.). For clusters with secondary peaks, we assumed that the peaks
are separable when their overlapping is $\le 20 \%$ and their distance
is $\ge$ 1000 \ks.

Out of 172 clusters, we found six clusters appearing as two separable
peaks and 17 which are not perfectly separable according
to our definition. These 17 clusters are discussed in \S~4. 
In the following analyses we consider the remaining 155 clusters.

The combination of position and velocity information, represented by
plots of velocity vs. clustercentric distance, can reveal the presence
of surviving interlopers (see e.g. Kent \& Gunn 1982; Reg\"os \& Geller
1989 and Figure~1).  We identified them by applying the fixed gap
method to a bin shifting along the distance from the cluster
center. We used a gap of 1000 \ks and a bin of 0.4 \h, or a larger bin
in order to have at least 15 galaxies. When the whole cluster was
analyzed, we removed the interlopers. We iterated the procedure until
the number of cluster members was stable. Hereafter we refer to this
procedure as "shifting gapper". Our procedure has the advantage of
being independent of hypotheses about the poorly known dynamical
status of the cluster, while the procedure used by M96 is based on physical
assumptions about the cluster mass profile.

Table~2 lists the 155 clusters which do not show any ambiguous
situation of peak overlapping. In Col.~(1) we list the cluster names;
in Col.~(2) the number of galaxies in each peak found by the adaptive
kernel method; in Col.~(3) the final number of galaxies used to
compute the mean (galactocentric) redshift [ Col.~(4)], and the $\sigma$
with bootstrap errors at the 68\% c.l.[ Col.~(5)], for each peak.

In the above selection of cluster membership, in order to maximize the
information available, we have used all the galaxies in our
sample. Since the value of $\sigma$, used in the following analyses,
depends both on galaxy velocities and on the spatial distribution, it
would be better to have samples complete up to a limiting
magnitude. Therefore, we have rejected the faintest galaxies of a few
clusters so as to eliminate an apparent trend of the limiting
magnitude to vary with the clustercentric distance.

Moreover, we did not consider galaxies beyond 5 \h from the cluster center.

\subsection{The Effect of Late-Type Galaxies}

For several clusters considered, there is only  partial information regarding
galaxy morphology.  For 40 clusters we have both a sufficient number of
early-type galaxies (ellipticals and lenticulars) and of late-type
galaxies (spirals and irregulars).  We checked for different means and
variances in the velocity distribution of these galaxy populations by
applying the standard means- and F-test (Press et al. 1992, see also
G96).  In order to consider the possible variation of $\sigma$ with
distance from the cluster center, in the above tests  we
considered only the largest area occupied by both galaxy populations.
We found that 12 of the 40 clusters show evidence of kinematical
differences (at the 95\% c.l.).

For several other clusters, in particular ENACS clusters, we have at
least some information regarding spectral features, e.g. the presence of
emission lines, A-type spectra, and starburst spectra, which can be
taken as indications of a late morphological type.  In particular, the
analysis of ENACS data shows that emission-line galaxies are
generally spirals and that their $\sigma$ within a cluster is larger
than the $\sigma$ of the other galaxies of the cluster (Biviano et
al. 1996). Therefore, we repeated the above analysis, considering also
spectral information for galaxies whose morphological type was unknown. All
emission-line (A-type spectra, starburst spectra) galaxies were
treated as late-type galaxies and all galaxies without these
spectral features were grouped together with early-type galaxies.  We
found other eight clusters showing a  kinematical difference between the two
galaxy populations.

Table~3 lists the 20 clusters out of the 79 analyzed that show significant
kinematical differences. In Col.~(1) we list the cluster names; in
Cols.~(2) and (3) the number of early- and late-type galaxies used for
each cluster; in Cols.~(4) and (5) $P_m$ and $P_F$, the probabilities
that mean values and velocity dispersions of early- and late-type
galaxy velocity distributions may be different, according to the mean-
and F-tests, respectively; in Col.~(6) the relevant morphology
reference sources.

The kinematical difference between early- and late-type galaxy
populations may be induced either by the presence of infall of spirals
on the cluster, or by spiral-rich substructures, or by some remaining
interlopers. By supposing that cluster dynamics is better represented
by early-type galaxies, we considered only this galaxy population when
we found significant kinematical differences (at the 95\% c.l.).

Here we are interested in determining the effect of neglecting
morphological/spectral information. Of the 79 clusters analyzed, only
14 show a difference in $\sigma$. For these clusters, the $\sigma$
value computed by using the global population is larger by 87$\pm18$
\ks, on average, with respect to the $\sigma$ value computed by using
only early-type galaxies. This difference is lower than that estimated
by G96 in their sample, maybe because of the better interloper
rejection, and it is roughly comparable within error estimates (see
Table~2).  For another 58 clusters, we have at least a sufficient number
of early-type galaxies to compute $\sigma$, which is, on average, 14
\ks lower than $\sigma$ computed for the global population. Thus our
analysis, which deals with about $80\%$ of the cluster sample, shows
that the morphological effect, although possibly relevant for a few
clusters, is negligible for $\sigma$-distribution.

\includegraphics{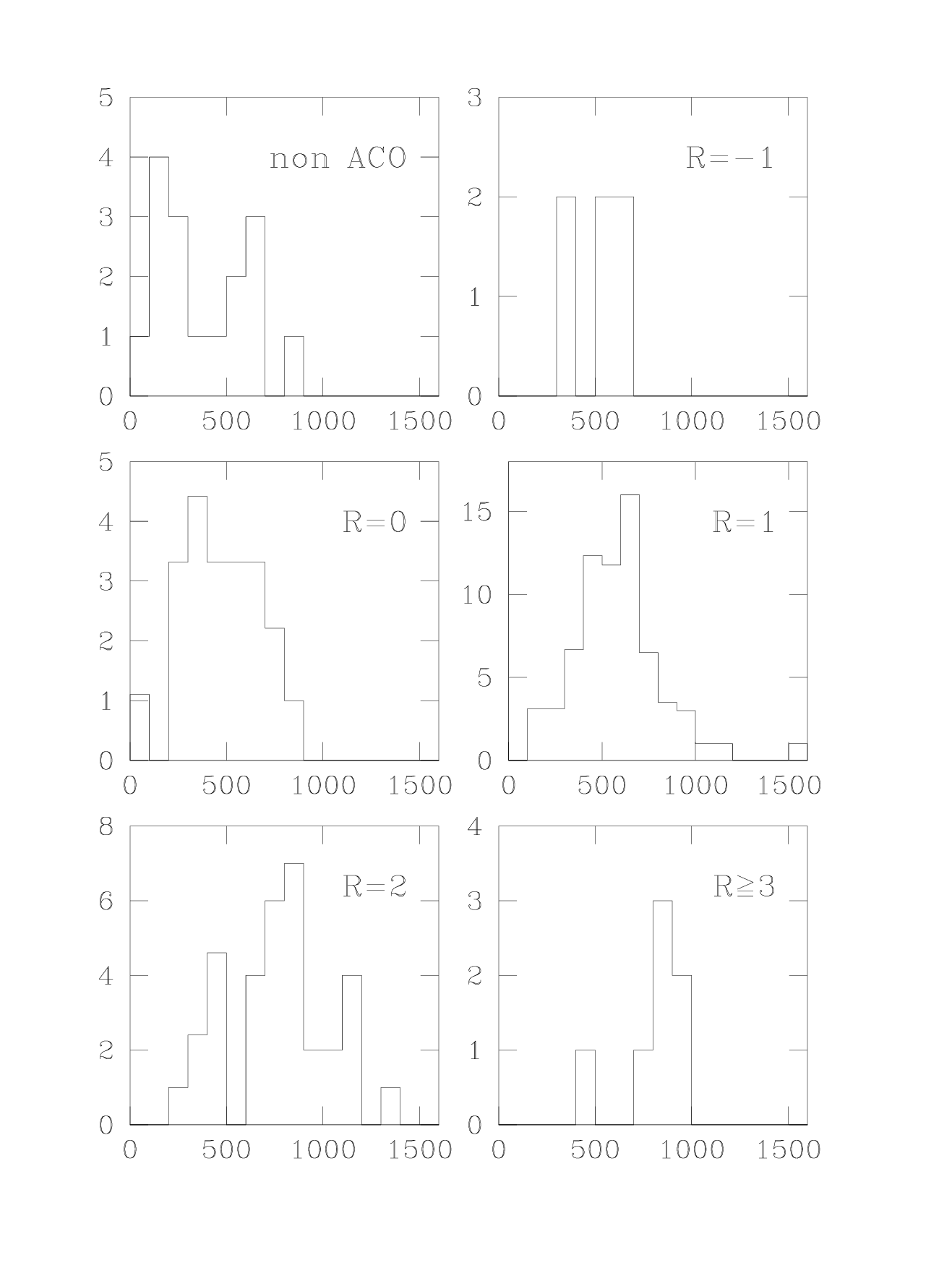}
\vspace{13.5cm}
\vspace{1mm}
{\small\parindent=-3.5mm
{\rm Fig.}~3.---The $\sigma$-distributions for non Abell-ACO
clusters and for clusters belonging to different Abell richness
classes $R$.  The velocity dispersions on the x-axis are in \ks.
}

\section{VDP and $\sigma$ Estimates}

We checked the presence of velocity gradients in the cluster velocity
field, performing a multiple linear regression between the galaxy
velocities and positions, and we evaluated their significance as in
G96. These velocity gradients may be produced by asymmetrical effects,
e.g. the presence of internal substructures, possible cluster
rotation, the  presence of other structures on larger scales such as nearby
clusters, surrounding superclusters, filaments.  For the 27 clusters
having significant gradients, we applied a correction by subtracting
the velocity gradient from each galaxy velocity and renormalizing the
velocities so as to leave their average velocity unchanged.  This
correction results in an a mean decrease in $\sigma$ of 74$\pm10$
\ks.

Finally, we considered the VDP, which, at a given radius is the average
l.o.s. velocity dispersion within this radius, i.e. it is evaluated
by using the velocities of all the galaxies within this radius (see
Figure~2). The VDPs may present different behaviours in the central
regions ($\lesssim$ 1 \h), which may probably depend also on the
choice of cluster center. However, most VDPs become flat in the
external cluster regions. This suggests that the final value of VDP is
representative of the total kinetic energy of galaxies. So we adopt
the final value of VDP, i.e. the velocity dipersion computed by using all
the cluster galaxies, as the value of $\sigma$ except for the
following cases.

Four clusters (A2440, A3225, A3691, A3764) show VDPs which appear very
far from flatness in the external sampled region, but they show a
sharp decrease, so we regard the final value of $\sigma$ as beeing
only an upper limit.

The VDPs of clusters A3391 and A4067, after the usual decreasing
behaviour, show an increase in the external region. Similarly, the
VDPs of clusters A1185 and AWM1 show a slowly increasing trend in the
external region.  This anomalous VDP increase in the external cluster
region can be explained by the presence of galaxies, which survived our
rejection procedure, belonging to a nearby cluster or to the field. In
fact, the VDP of cluster A3391 is shown to be affected by the
presence of a nearby cluster (see G96). For these four clusters we
adopted the value of $\sigma$ obtaining before the increase in VDPs towards
external regions.

The VDP of the very rich cluster A2255 is particularly anomalous since
it always increases, possibly becoming flat only at about 1.5-2\h, well
beyond the usual region (see e.g. the cluster A2063). The scarcity
of data does not allow us to be more precise regarding this cluster, so we
prefer to reject it from our list.

The $\sigma$ values for our  clusters are presented in Table~2.

M96 found no clusters with $\sigma >1200$ \ks, while in our sample we
found two clusters with very high $\sigma$ (A2256, A2319), which are
not present in the M96 sample. M96 claimed that their
cluster-membership selection procedure is decisive in reducing the
estimate of $\sigma$; in particular they compare their procedure with
that used by Z93. We therefore checked our procedure against that of
M96, comparing $\sigma$ for the galaxy systems that the two samples
have in common. Among these, eight clusters are recognized as beeing
multiple peaks (in redshift or spatially) by our procedure and so our
$\sigma$ values are clearly lower than M96 values. For the other 66
systems in common, there is good agreement ($<\sigma_{M96} -
\sigma_{our}>=9\pm 7$ \ks) between the $\sigma$ of M96 and ours,
computed after peak selection and the "shifting gapper"
procedure. Our further analyses generally reduce the adopted value of
$\sigma$, so that $<\sigma_{M96} - \sigma_{our}>=41\pm 10$ \ks. This
comparison suggests that our high $\sigma$ values are not induced by
our procedure.  However, the value of $\sigma=1545$ \ks for cluster
A2319 is anomalous for its Abell class, $R=1$, lying beyond 4 standard
deviations from the mean (see Table~6). So we exclude it in the
computation of the $\sigma$-distribution, although it would not cause
any significant variation in the high$-\sigma$ tail of the distribution.

\includegraphics{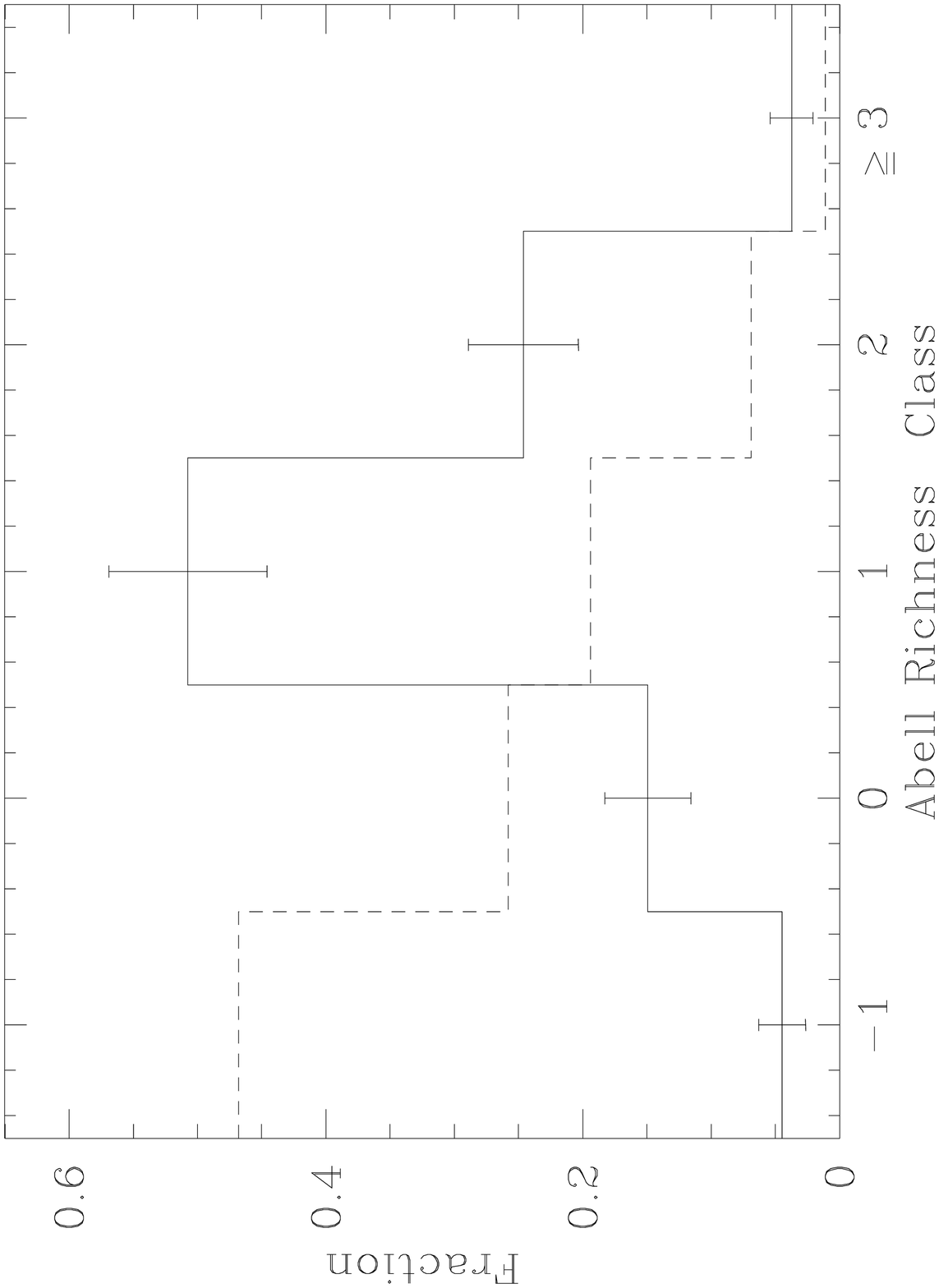}
\vspace{7.5cm}
\vspace{1mm}
{\small\parindent=-3.5mm
{\rm Fig.}~4.---The Abell richness class distribution for our
sample (solid line) and for the EDCC catalog (dashed line).  The
$R$-distribution of the EDCC catalog was obtained by re-scaling the
EDCC richness to the Abell-ACO ones (see also Biviano et al. 1993).}

\section{Cluster Substructures and Multipeaked Clusters}

Although the existence of cluster substructures is well established,
it is not yet well understood to what extent it influences cluster
dynamics (e.g. Fitchett 1988; Gonz\'alez-Casado, Mamon, \&
Salvador-Sol\'e 1994; West 1994).  Here we take into account the
possible presence of cluster substructures which can strongly modify
the estimate of $\sigma$.

The strongly overlapping peaks presented by 17 of our clusters in
their velocity distribution may have several explanations. The peaks
may be different systems superimposed along the line of sight. For
instance, Lucey, Currie, \& Dickens (1986) found that at least a
minority of galaxies in the secondary peak in cluster A3526 is actually
distant from the primary peak. Otherwise, the peaks we find could
indicate the presence of substructures in a single system and, in this
case, it is uncertain whether we should choose between the $\sigma$ of the
substructures and the $\sigma$ of the whole system as an indicator of
the total cluster potential.  Thus, we prefer to treat these multipeaked
clusters in two ways: both disjoining and not disjoining the peaks, in order to
obtain two boundary limits for the true $\sigma$-distribution.

In order to compute $\sigma$ for the 17 clusters, we repeated all the
above analyses relative to galaxy morphology and velocity
gradients. Since the dynamics of these clusters is probably more
troublesome than usual, we simply adopted the value of $\sigma$
computed for the whole galaxy sample. In Tables~4 and 5 we list the
17 clusters which show an ambiguous situation of peak overlapping,
considering the peaks as joined and disjoined, respectively.  In
Col.~(1) we list the cluster names; in Col.~(2) the number of galaxies
in each peak found by the adaptive kernel method; in Col.~(3) the final
number of galaxies used to compute the mean (galactocentric) redshift
[Col.~(4)], and the $\sigma$ with bootstrap errors at the 68\% c.l.
[Col.~(5)], for each peak.

\section{The $\sigma$-Distribution}

Our cluster sample, as it spans a larger range of richness than other
studies, allows us to better analyze the bias introduced by the $R$
completeness limit.  We exclude from this analysis the multipeaked
clusters.

The non Abell-ACO clusters span a large range of $\sigma$ (see
Figure~3), confirming that the Abell-ACO catalog has not detected some rich
clusters.  This fact was widely discussed by some authors
(e.g. Scaramella et al.  1991) who pointed out that the  Abell-ACO catalog is
incomplete. In particular, Lumsden et al. (1992) showed that the ACO
incompleteness is larger for poorer clusters. For this reason, we
prefer to use the richness class distribution of the Edinburgh-Durham
Cluster Catalog (EDCC, Lumsden et al. 1992), rather than that of Abell-ACO.

Table~6 lists the average value of $\sigma$ for each $R$.  With the
present data, the $R=-1$ class does not appear to be different from
the $R=0$ class. The average $\sigma$ significantly increases with $R$
in the range $R=0-2$, but hints at a possible flattening of this
relation for $R\ge3$ (see also Danese et al. 1980; Girardi et
al. 1993). Moreover, there is a large overlapping in $\sigma$ values
for clusters of different $R$ (see Figure~3).  For instance, in the
range 800-900 \ks we found that the ratio of the number of clusters
with $R=0$ to the number of these with $R\ge1$ is about 23\%,
weighting such numbers by the class frequencies of the EDCC
catalog. This ratio increases to 43\% in the range 700-800 \ks.  This
suggests a considerable incompleteness already for $\sigma < 800-900 $
\ks, when the cluster sample is only complete down to $R=1$.

In order to extend the $\sigma$ completeness of the
$\sigma$-distribution, one must consider also $R\le0$ clusters.  Our
aim is to obtain a cluster sample representative of the nearby
Universe and complete for poor clusters.

Unfortunately, our $R$-distribution is biased towards richer clusters
with respect to the EDCC catalog (see Figure~4), which is claimed to
be complete also for very poor clusters (Lumsden et al. 1992).  We
obtained a more representative cluster sample by resampling our
clusters to mimic an universal $R$-distribution (see also Girardi et
al. 1993; Biviano et al. 1993).  We computed the $\sigma$-distribution
by using 10000 random extractions of observed $\sigma$, distributed
according to the EDCC $R$-distribution.  In order to take into account
the upper limited values, we redistributed the upper limits among the
lower detected values.  In the case of clusters with $n$ peaks, the
value of $\sigma$ of each peak is weighted by $1/n$.  The same
weighting was also used to compute the $\sigma$ averages presented in
Table~6 and to obtain the histograms in Figure~3.

We applied the resampling procedure to the 148 Abell-ACO $R\ge0$
clusters.  The ambiguity of clusters with uncertain dynamics (\S~4)
results in two $\sigma$-distributions. These cumulative
$\sigma$-distributions and the one obtained by weighting the two cases
in the same way are plotted in Figure~5. The upper and lower
cumulative distributions are so close that they lie well within the
error bands of the intermediate one.  Therefore, in the following
analyses we adopt the intermediate distribution as the
$\sigma$-distribution complete for $R\ge0$ clusters.

\includegraphics{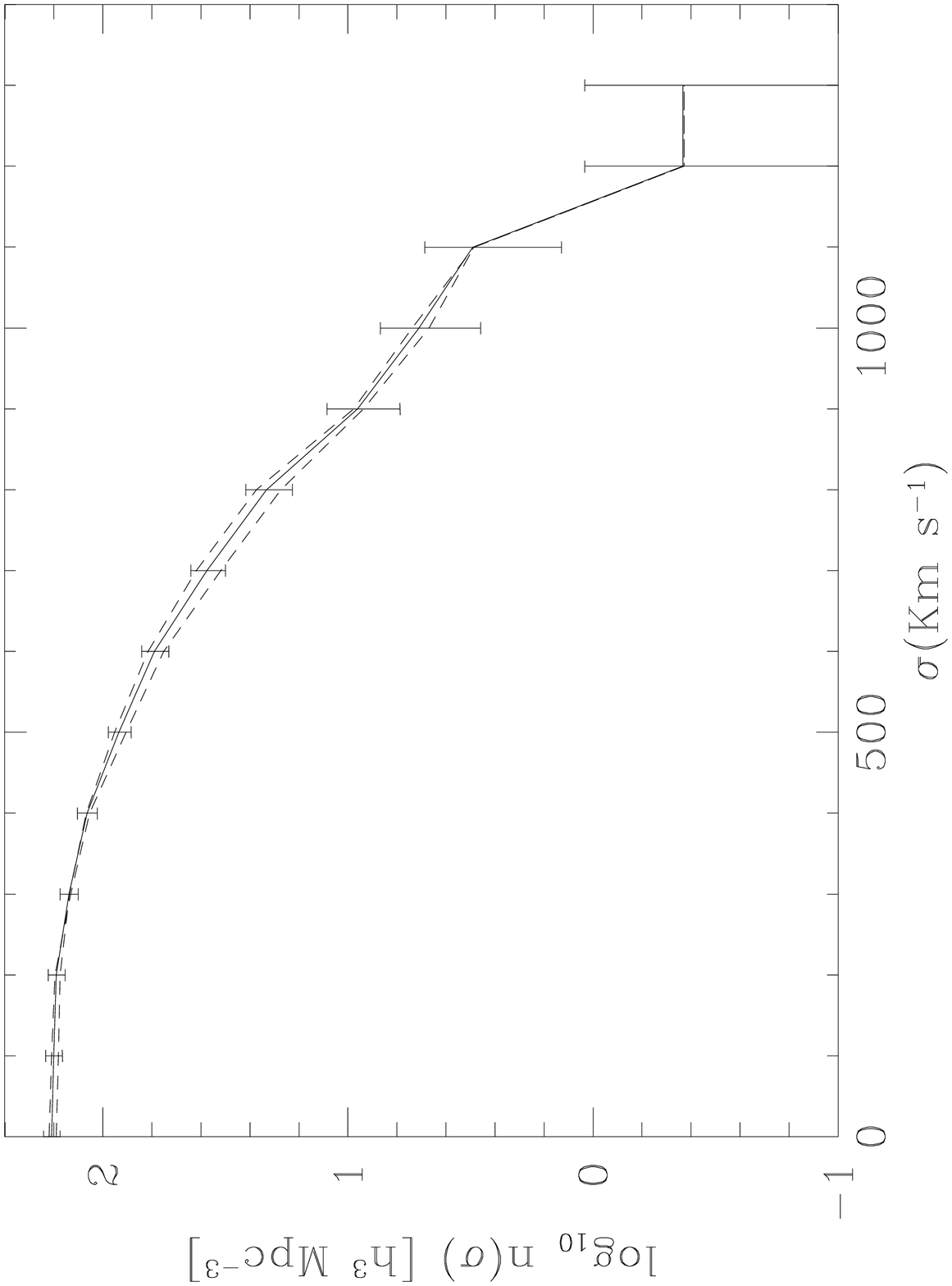}
\vspace{7.5cm}
\vspace{1mm}
{\small\parindent=-3.5mm
{\rm Fig.}~5.---The cumulative $\sigma$-distributions for $R
\ge 0$ clusters.  The dashed lines consider multipeaked clusters as
treated in Tables~4 and~5. The intermediate case (solid line) is
represented with its error bars.}
\vspace{5mm}

In order to further extend the completeness of our
$\sigma$-distribution, we considered all the 153 clusters with
$R\ge-1$. Since we found no difference in mean $\sigma$ between $R=0$
and $R=-1$ clusters, and we had a small number of $R=-1$ clusters, we
treated these two classes together. We again applied the above
procedure, now obtaining a $\sigma$-distribution complete for $R\ge-1$
clusters.

\includegraphics{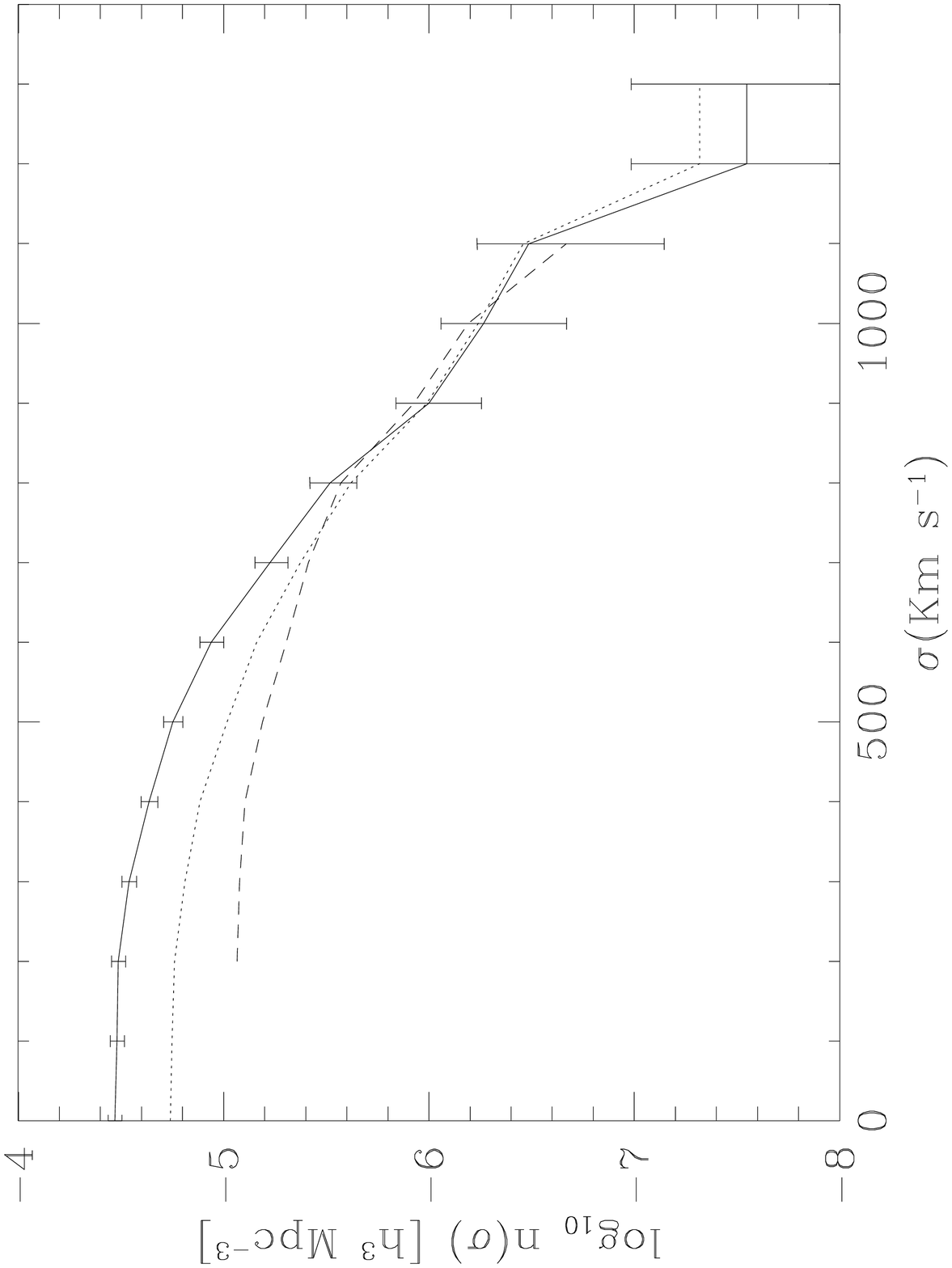}
\vspace{7.5cm}
\vspace{1mm}
{\small\parindent=-3.5mm
{\rm Fig.}~6.---The comparison of cumulative
$\sigma$-distributions: our results for $R\ge -1$ clusters (solid
line) and for $R\ge 0$ clusters (dotted line); M96 for clusters with
$R \ge 1$, (dashed line).}
\vspace{5mm}

To normalize our distributions we adopted the cluster volume density
$8.6\cdot 10^{-6}$\h$^{-3}$ for clusters with $R\ge1$ (M96) 
scaled to the EDCC class frequencies.
The M96 value was corrected for the incompleteness of the Abell-ACO
catalog with respect to the EDCC catalog, so that our normalization is
consistent with our procedure.

Figure~6, which compares our cumulative distribution with that of M96,
shows how the better $R$-completeness of our  cluster sample
improves the sample completeness with respect to $\sigma$.  The
Kolmogorov-Smirnov test assures that our $\sigma$-distribution,
complete for $R\ge0$ ($R\ge-1$) clusters, is not significantly
different from the less $R$-complete ones, presented in recent works
(Girardi et al. 1993; Z93; Collins et al. 1995; M96), within their
supposed completeness limit ($\sigma\sim$ 800 \ks).  The estimate of
the (better) completeness limit of our $\sigma$ distribution requires
other analyses.

\includegraphics{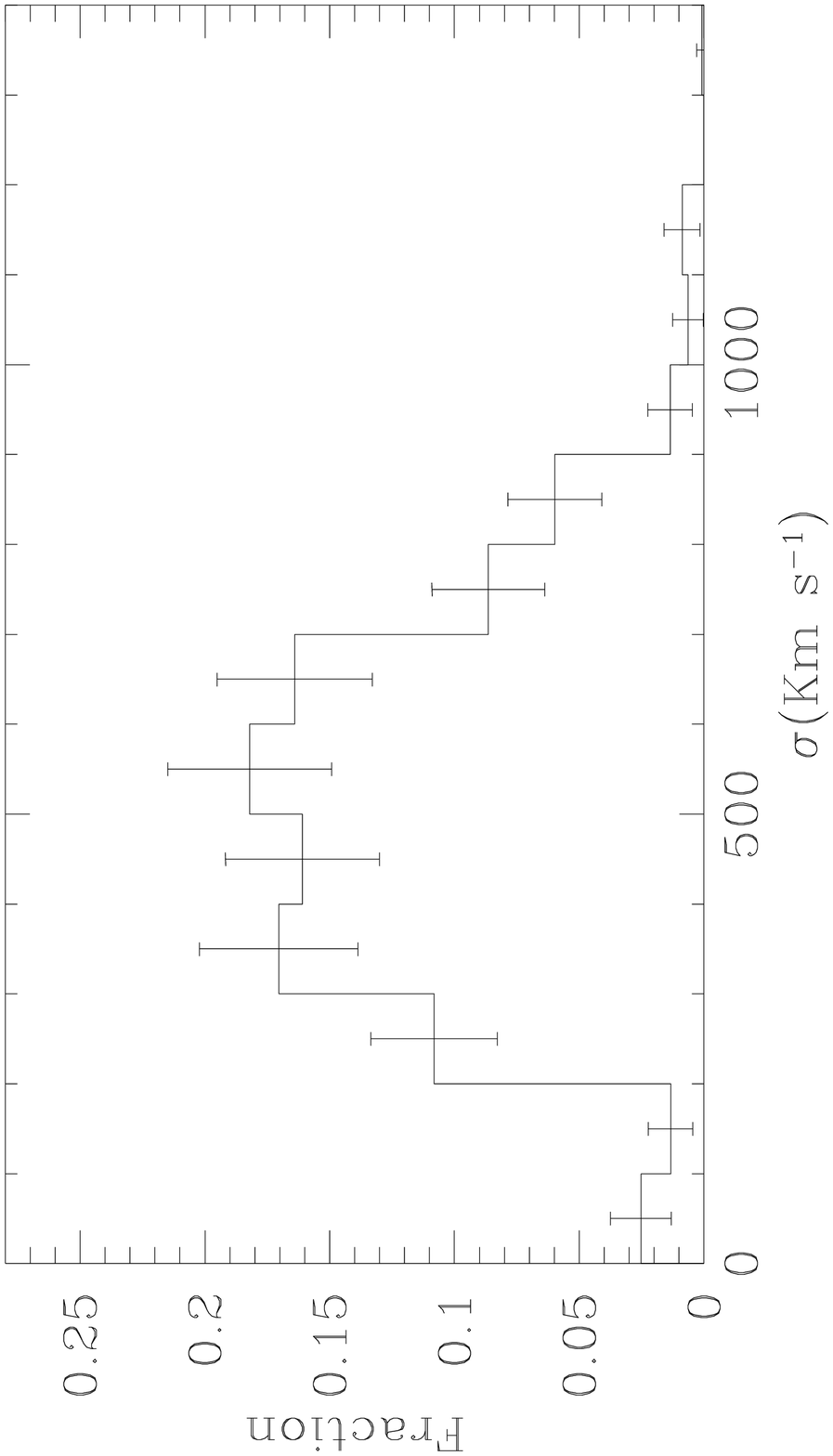}
\vspace{6cm}
\vspace{1mm}
{\small\parindent=-3.5mm
{\rm Fig.}~7.---The $\sigma$-distribution of clusters with $R\ge -1$.}
\\

\includegraphics{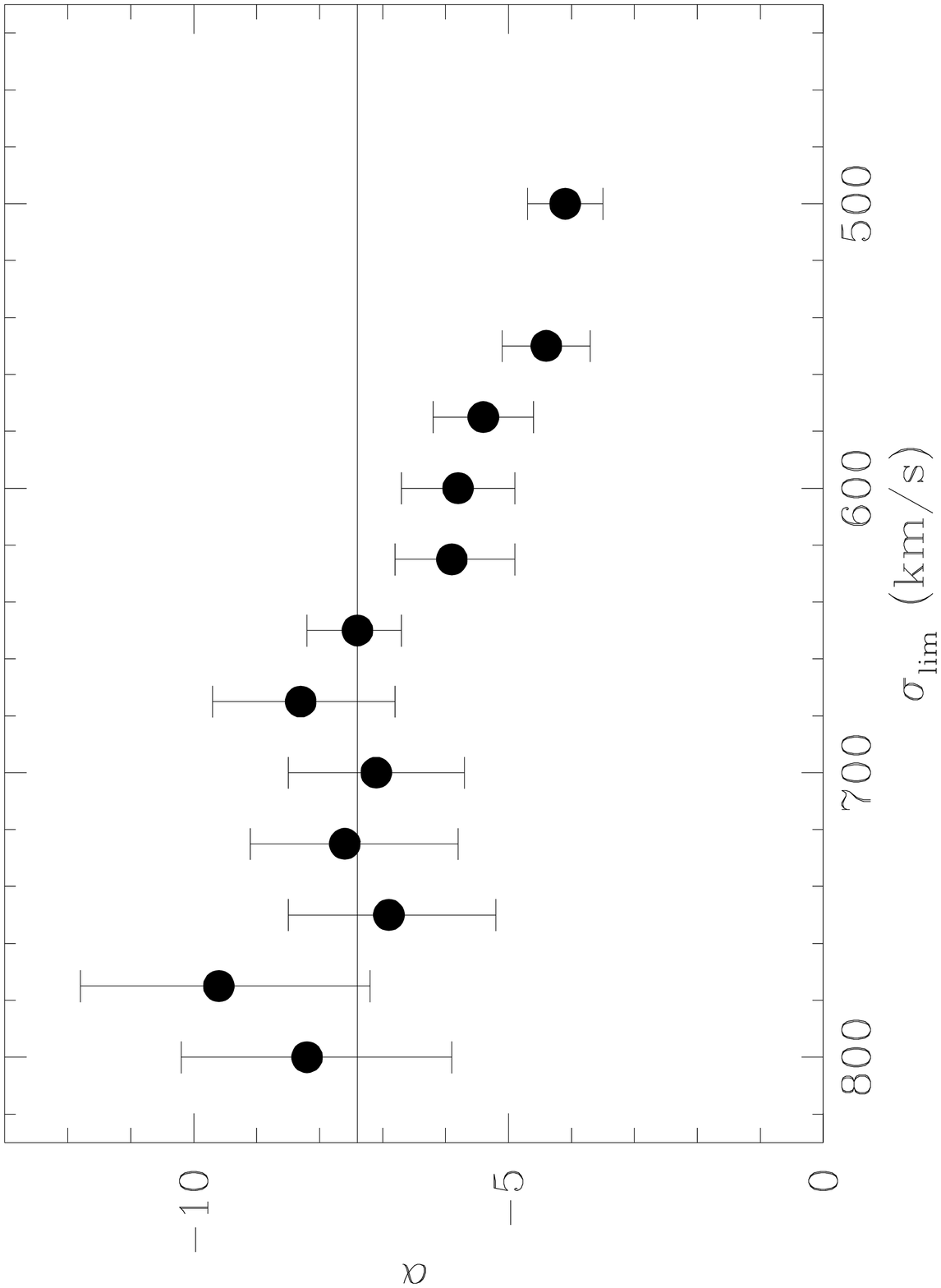}
\vspace{7.5cm}
\vspace{1mm}
{\small\parindent=-3.5mm
{\rm Fig.}~8.---The exponents $\alpha$ of the fitted power law
obtained from fitting in larger ad larger ranges ($\sigma \ge
\sigma_{lim}$).  The horizontal line shows the $\alpha$ value adopted
in this paper.}
\vspace{5mm}

The inspection of our $\sigma$-distribution (Figure~7) shows an
obvious incompleteness below $\sigma\sim 500$ \ks.  We fitted our data
to a power law ($dN\propto \sigma^{\alpha}d\sigma$) by using the
Maximum Likelihood method.  We made this fit in different ranges by
decreasing the supposed completeness limit from 800 \ks to 500 \ks.
The exponent $\alpha$ appears to be roughly stable, within the errors,
down to a limiting value of about $650$ \ks (Figure~8).  Consequently, if
we assume that the $\sigma$-distribution is well described by a power
law, this value of $\sigma$ represents our completeness limit.  The
fit in this range, acceptable according to the Kolmogorov-Smirnov
test, gives $\alpha=-(7.4^{+0.7}_{-0.8})$.

\includegraphics{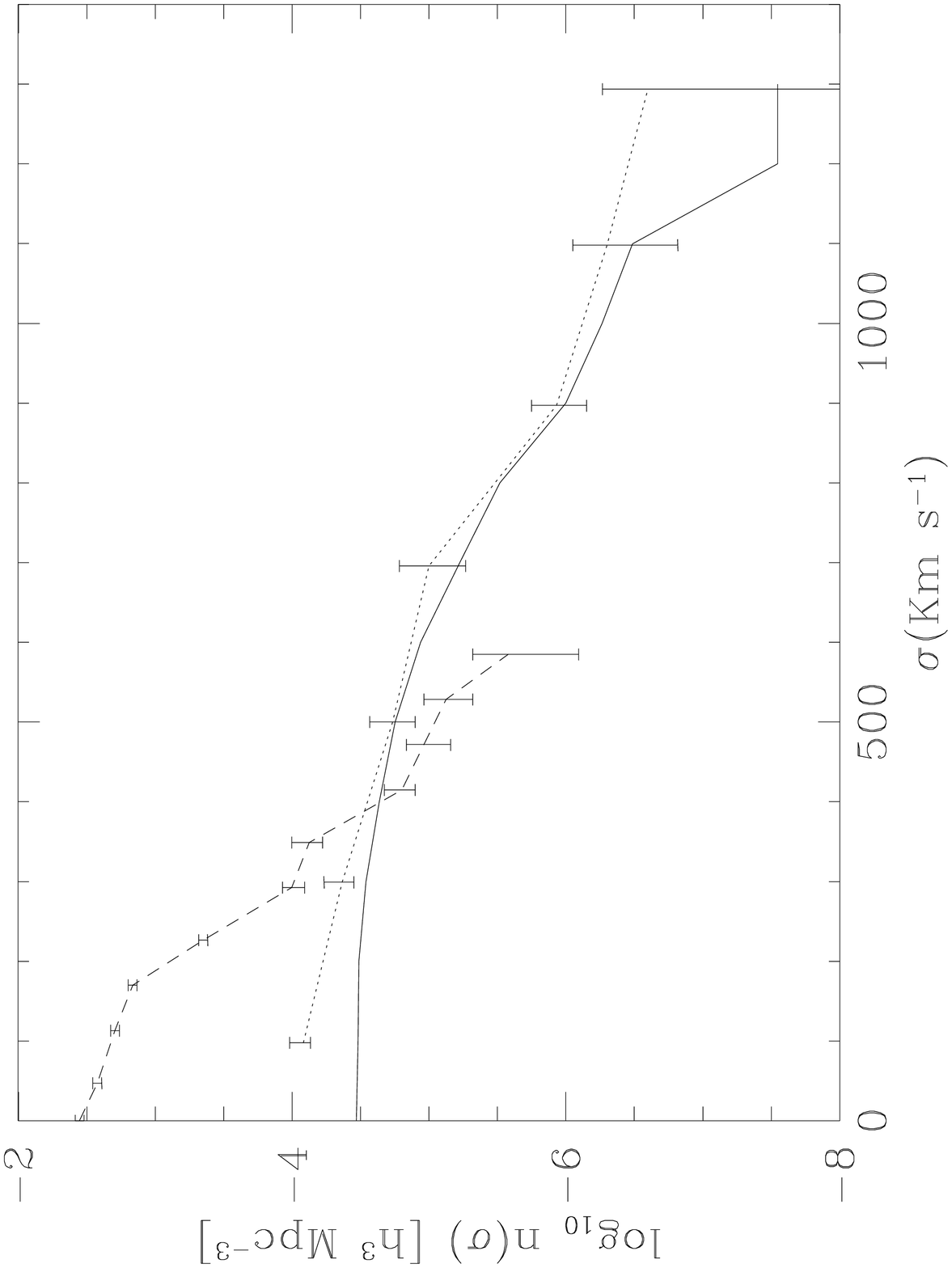}
\vspace{7.5cm}
\vspace{1mm}
{\small\parindent=-3.5mm
{\rm Fig.}~9.---The comparison of cumulative
$\sigma$-distributions: our results for $R\ge -1$ clusters (solid
line); Zabludoff et al. (1993a) for groups and clusters (dotted line);
Moore et al. (1993) for groups (dashed line).}

\section{Discussion and Conclusions}

The $\sigma$-values we present here were obtained by taking into
account the effects which could modify the final
$\sigma$-distribution. In particular, our $\sigma$ values are
independent of possible velocity anisotropy in galaxy orbits. The
distribution of spatial $\sigma$ is recovered by applying the usual
projection factor ($\sqrt{3}$) which relates the line-of-sight to the
spatial $\sigma$. Also in the presence of cluster asphericity, the
projection factor, averaged over a sample of randomly oriented
clusters, is still the same (see G96), so that possible errors cancel
each other out in the $\sigma$-distribution.

At present, an extended volume-complete sample of clusters (with known
$\sigma$) which includes poor clusters is not available.  On the
other hand, one needs to take into account also poor clusters in order to
obtain a better $\sigma$ completeness.  By resampling 153 Abell-ACO
clusters, according to the richness class frequencies of the EDCC
catalog, we obtain a cluster sample representative of the nearby
Universe down to and including poor clusters.

Our $\sigma$-distribution agrees with previous optical results and
extends the $\sigma$ completeness limit down to at least 650 \ks.  In
this range, a fit of the form $dN\propto \sigma^{\alpha}d\sigma$ gives
$\alpha=-(7.4^{+0.7}_{-0.8})$.

 This result agrees with the distribution functions of X-ray
temperatures.  By converting the X-ray temperature $T$ to the
equivalent $\sigma$ according to the relation $T\propto \sigma^2$, one
obtains the values of $\alpha=-(8.86\pm0.74)$ and
$\alpha=-(8.4\pm1.0)$, by using the results of Edge et al. (1990) and
Henry \& Arnaud (1991), respectively.  If one assumes the empirical
relation $T\propto \sigma^{1.64}$ obtained by G96, the agreement is
even better: $\alpha=-(7.45\pm0.61)$ and $\alpha=-(7.0\pm0.8)$,
respectively.

In order to more fully verify the completeness of our distribution in the
low $\sigma$ range, we also considered the results for galaxy groups,
as the boundary between groups and poor clusters is very fuzzy. We
compared our cumulative distribution with the one obtained by Z93, who
considered groups and clusters, and with that of Moore et al. (1993)
who considered groups, reporting their exact cluster
volume-densities (Figure~9). Considering the error bands, we agree
with the  Z93 distribution down to 400 \ks, well beyond our supposed
completeness limit.  The comparison with the $\sigma$-distribution of
Moore et al. (1993) suggests that their volume-density of groups for
$\sigma\ge 400$ \ks may be underestimated with respect to the
volume-density of our poor systems.  Unfortunately, we cannot draw any
firm conclusions about our possible incompleteness level, and hence
about the behaviour of the $\sigma$-distribution, for $\sigma<650$ \ks,
because the $\sigma$-distribution of galaxy groups is still poorly
known.  In particular, taking the groups identified by Garcia (1993)
and analyzed by Giaiotti et al. (1996), we verified that the shape of the
$\sigma$-distributions strongly depends on the choice of the selection
method (hierarchical, friends of friends, and a combination of the two
methods, see Garcia 1993).

For the range of $\sigma$-distribution pertaining to clusters,
there is good agreement among the results obtained by different
authors. In particular, the improved statistics used in this work allow us
to further reduce observational errors. Therefore, our
distribution is already sufficiently extended and precise to be a good
constraint for theories of large-scale structures formation.  The
comparison with theoretical distributions (e.g. Jing \& Fang 1994;
Crone \& Geller 1995) allows us to reject some models, e.g. the
standard cold dark matter model. However, the new COBE normalization
(G\'orski et al. 1995) might further reduce the range of acceptable
models.  We have not made any comparison with theoretical mass functions, since
the mass-$\sigma$ relation is still poorly known.  For instance, in
our case, the virial theorem relates mass and dispersion only after
an assumption regarding the relative distribution of mass and galaxies.

Since observational $\sigma$-distributions, in spite of their
uncertainties, are more reliable than mass-distributions, we
stress the need for a large computational effort in order to obtain good
$\sigma$-distributions from N-body simulations for a variety of
cosmological models.

\acknowledgments 

We are particularly indebted to the ENACS team for having kindly
provided us with their data, based on observations collected at the
European Southern Observatory (ESO), in advance of publication.

We thank Dario Giaiotti, who provided us with his results on galaxy
groups, Andrea Biviano, Stefano Borgani, Yi-Peng Jing, Nicola Menci,
Manolis Plionis, Massimo Ramella, and Riccardo Valdarnini for useful
discussions. 

We thank the referee David Merritt for his useful suggestions.  Some
misprints in the tabulated redshifts were noted by the first anonymous
referee.

This work was partially supported by the {\em Ministero per
l'Universit\`a e per la Ricerca scientifica e tecnologica}, by the
{\em Italian Space Agency} (A.S.I.), and by the {\em Consiglio
Nazionale delle Ricerche (CNR-GNA)}.

\end{multicols}

\newpage

\small
\input{tab/tab1.tex}
\input{tab/tab2.tex}

\input{tab/tab3.tex}
\input{tab/tab4.tex}
\input{tab/tab5.tex}

\input{tab/tab6.tex}
\end{document}

%% file: tab/tab1.tex
\begin{deluxetable}{lrcl|lrcl}
\tablecolumns{8}
\tablewidth{0pc}
\tablecaption{The Cluster Sample}
\tablehead{
\colhead{Name} &
\colhead{N}    & 
\colhead{$R$}  & 
\multicolumn{1}{c|}{References}&
\colhead{Name} &
\colhead{N}    & 
\colhead{$R$}  & 
\colhead{References}\\
\colhead{(1)} &
\colhead{(2)}    & 
\colhead{(3)}  & 
\multicolumn{1}{c|}{(4)}&
\colhead{(1)} &
\colhead{(2)}    & 
\colhead{(3)}  & 
\colhead{(4)}
}
\startdata
A13         & 44& 2& E &                       A2142       &119& 2& Oeg110,1;Hin74,116\nl      
A85         &185& 1& Bee102,1581;Mal104,495&   A2151 Hercules     &105& 2& Dre95,284\nl             
A87         & 42& 1& E&                        A2197         & 45& 1& Gre286,422\nl                 
A118        & 38& 1& E&                        A2199         & 71& 2& Gre286,422\nl                 
A119        &142& 1& E;Fab105,788&             A2255         & 31& 2& Sta228,379\nl                 
A151        &142& 1& E;Pro258,243&             A2256         & 89& 2& Fab336,77\nl                  
A168        &111& 2& E;Fab82,187;Zab106,1273&  A2319         &128& 1& Oeg110,1\nl                   
A193        & 65& 1& Hil106,831&               A2353         & 31& 1& E\nl                          
A194        &267& 0& Cha95,999&                A2362         & 33& 1& E\nl                          
A229        & 39& 1& E&                        A2401         & 30& 1& E\nl                          
A256        & 36& 1& Zab106,1273&              A2426         & 36& 2& E\nl                          
A262        & 88& 0& Gio262,442;Gre243,411&    A2440         & 30& 0& Bee102,1581\nl                
A295        & 65& 1& E;Huc41,31;Zab106,1273&   A2500         & 36& 1& E\nl                          
A367        & 30& 1& E&                        A2538         & 45& 1& Col224,453\nl                 
A399        & 90& 1& Hil106,831&               A2554         & 41& 3& Col224,453\nl                 
A400        &109& 1& Bee400,410&               A2569         & 41& 1& E\nl                          
A401        &117& 2& Hil106,831&               A2589         & 33& 0& Bee102,1581\nl                
A420        & 33& 1& E&                        A2593         & 37& 0& Bee102,1581\nl                
A426 Perseus&200& 2& Ken88,697&                A2634         &315& 1& Sco444,41;Hin85,626;\nl       
A458        & 45& 2& Col224,453&                      &   &  & Pin416,36; Zab74,1;\nl               
A496        &166& 1& Mal104,495;Qui100,1424&                 &   &  & Bot335,617;But57,665\nl       
A514        &111& 1& E&                        A2644         & 35& 1& E\nl                          
A524        & 43& 1& E&                        A2666         & 27& 0& Sco444,41\nl                  
A539        &289& 1& Ost96,1775&               A2670         &303& 3& Sha231,479\nl                 
A548        &134& 1& Dre95,284&                A2715         & 34& 2& E\nl                          
A569        & 41& 0& Bee102,1581&              A2717         & 81& 1& E;Col224,453\nl               
A576        & 51& 1& Hin87,1656&               A2721         &104& 3& Tea72,715;Col224,453\nl       
A634        & 38& 0& Ste20,478&                A2734         &116& 1& E\nl                          
A754        & 89& 2& Dre95,284&                A2755         & 36& 2& E\nl                          
A957        & 54& 1& E;Bee102,1581&            A2798         & 31& 1& Col274,1071\nl                
A978        & 73& 1& E&                        A2799         & 42& 1& E\nl                          
A999        & 45& 0& Cha94,571&                A2800         & 46& 1& E\nl                          
A1016       & 44& 0& Cha94,571&                A2819         &124& 2& E\nl                          
A1060 Hydra &177& 1& Ric67,237;Ric77,237&      A2854         & 35& 1& E\nl                          
A1069       & 40& 0& E&                        A2877         &110& 0& Mal104,495\nl                 
A1142       & 66& 0& Gel89,319&                A2911         & 44& 1& E\nl                          
A1146       & 84& 4& Tea72,715&                A3093         & 40& 2& E\nl                          
A1185       & 77& 1& Bee102,1581&              A3094         & 99& 2& E\nl                          
A1228       & 39& 1& Zab106,1273&              A3111         & 48& 1& E\nl                          
A1314       & 30& 0& Del107,427;Huc41,31&      A3112         &128& 2& E;Mat124,L13\nl               
A1367       & 94& 2& Gav320,96;Gre222,784;&    A3122         &119& 2& E\nl                          
            &   &  & Tif222,54;Dic174,47&      A3126         & 45& 1& Col224,453\nl                 
A1631       & 90& 0& Dre95,284&                A3128         &222& 3& E;Col224,453\nl               
A1644       &102& 1& Dre95,284&                A3142         & 38& 1& E\nl                          
A1651       & 62& 1& Del107,427;Huc41,31&      A3151         & 43& 1& E\nl                          
A1656 Coma  &414& 2& Ken87,945&                A3158         &145& 2& E;Chi96,106;Luc203,545\nl     
A1736       &104& 0& Dre95,284&                A3194         & 33& 2& E\nl                          
A1775       & 77& 2& Oeg110,1&                 A3202         & 41& 1& E\nl                          
A1795       & 98& 2& Hil106,831&               A3223         &110& 2& E\nl                          
A1809       & 68& 1& E;Hil106,831&             A3225         & 44& 0& Col224,453\nl                 
A1983       &100& 1& Dre95,284&                A3266         &172& 2& Tea72,715\nl                  
A1991       & 25& 1& Bee102,1581&              A3334         & 41& 2& Col224,453\nl                 
A2029       & 93& 2& Oeg110,1&                 A3341         &118& 2& E\nl                          
A2040       & 67& 1& E;Zab106,1273&            A3354         &110& 1& E\nl                          
A2048       & 39& 1& E&                        A3360         & 40& 2& Col224,453\nl                 
A2052       & 60& 0& E;Qui90,410;Mal104,495&   A3376         & 84& 0& Dre95,284\nl                  
A2063       & 92& 1& Bee102,1581;Hil106,831&   A3381         & 64& 1& Dre95,284\nl                  
A2079       & 32& 1& Pos92,1238&               A3389          & 45& 0& Tea72,715\nl                 
A2092       & 30& 1& Pos92,1238&               A3391          & 65& 0& Tea72,715\nl                 
A2107       & 75& 1& Oeg104,2078&              A3395          &143& 1& Tea72,715\nl                 
A2124       & 67& 1& Hil106,831&               A3526 Centaurus&301& 0& Dic220,679;ESO\nl            
A3528          & 39& 1& E&                       A4038          & 99& 2& Ett276,689;Luc235,1177\nl 
A3532          & 44& 0& Cri179,108&              A4053          & 31& 1& E\nl                     
A3556       & 54& 0& Bar267,665&                 A4067          & 41& 1& Tea72,715\nl             
A3558 Shapley 8&398& 4& E;Bar267,665;Ric261,266;&S84            & 30&-1& Col274,1071\nl           
               &   &  & Tea72,715;Met225,581&    S373           & 57&-1& ESO\nl                       
A3559          & 69& 3& E&                       S463           &100&-1& Dre95,284\nl                 
A3562          &119& 2& E&                       S753           & 43&-1& Wil101,57\nl                 
A3571          & 72& 2& Qui101,475&              S805           &154&-1& Mal104,495\nl                
A3574          & 42& 0& Wil101,57&               S987           & 40&-1& Col274,1071\nl               
A3651          & 92& 1& E&                       AWM1           & 56&$-$& Bee283,33\nl                
A3667          &177& 2& E;Sod259,233&            AWM4           & 59&$-$& Del107,427;Huc41,31\nl      
A3691          & 36& 2& E&                       AWM7           & 33&$-$& Del107,427;Huc41,31\nl      
A3693          & 33& 1& E&                       C67            & 45&$-$& Col224,453\nl               
A3695          & 96& 2& E&                       CL2335-26      & 38&$-$& Sco444,41\nl                
A3703          & 32& 1& E&                       DC0003-50      & 55&$-$& Dre95,284\nl                
A3705          & 45& 2& Col224,453&              Eridanus       & 65&$-$& Wil98,1531\nl               
A3716          &138& 1& Dre95,284;Col224,453&    MKW1           & 39&$-$& Bee283,33\nl                
A3733          & 44& 1& E&                       MKW3S          & 30& 1 & Bee102,1581;Hil106,831\nl   
A3744          & 86& 1& E&                       MKW4S          & 39&$-$& Del107,427;Huc41,31\nl      
A3764          & 43& 1& E&                       MKW5           & 44&$-$& Del107,427;Huc41,31\nl      
A3806          &119& 2& E&                       MKW6A          & 37&$-$& Del107,427;Huc41,31\nl      
A3809          &127& 1& E&                       MKW10          & 69&$-$& Del107,427;Huc41,31\nl      
A3822          &101& 2& E&                       MKW12          & 66&$-$& Bee283,33\nl                
A3825          & 90& 1& E&                       Pegasus        & 78&$-$& Ric109,155;Bot57,423;\nl    
A3879          & 82& 2& E&                                         &   &   & Chi88,388\nl             
A3921          & 38& 2& E&                       Ursa           & 57&$-$& NGC\nl                    
A4008          & 43& 1& E&                       Virgo          &572&$-$& Bin90,1681\nl             
A4010          & 36& 1& E&                       &&&\nl                               
\tablecomments{The bibliographical references in Column 4 use a code consisting
of the first three letters of the last name of the first author, followed by
the journal volume number, and then, after a comma, the initial page number
of the paper. For example, the paper by Girardi \& al. 1993 in this code would
be ``Gir404,38''. The only exceptions are: E, Enacs data; ESO, the catalog
of Laubert \& al. 1989; NGC, the catalog of Tully 1988.}
\enddata
\end{deluxetable}

%% file: tab/tab2.tex
\begin{deluxetable}{lcccrr|lcccrr}
\tablecolumns{12}
\tablewidth{0pc}
\tablecaption{The Values of $\sigma$}
\tablehead{
\colhead{Name} &
\colhead{$N_p$}    & 
\colhead{$N_{mem}$}  & 
\colhead{z}&
\multicolumn{2}{c|}{$\sigma$ (km/s)} &
\colhead{Name} &
\colhead{$N_p$}    & 
\colhead{$N_{mem}$}  & 
\colhead{z}&
\multicolumn{2}{c}{$\sigma$ (km/s)} \\
\colhead{(1)} &
\colhead{(2)} & 
\colhead{(3)} & 
\colhead{(4)} &
\multicolumn{2}{c|}{(5)}&
\colhead{(1)} &
\colhead{(2)} & 
\colhead{(3)} & 
\colhead{(4)} &
\multicolumn{2}{c}{(5)}
}
\startdata
A85      & 130& 125& 0.0559&  969&$+$\phn 95/$-$\phn 61&A2319    & 127& 118& 0.0553& 1545&$+$\phn 95/$-$\phn 77\nl
A119     & 128&  62& 0.0438&  679&    $+$106/$-$\phn 80&A2353    &  24&  24& 0.1213&  597&$+$\phn 88/$-$\phn 66\nl                 
A151a    &  64&  64& 0.0537&  714&$+$\phn 73/$-$\phn 61&A2362    &  24&  24& 0.0616&  331&$+$\phn 68/$-$\phn 52\nl                 
A151b    &  29&   7& 0.0409&  385&    $+$118/$-$\phn 31&A2401    &  23&  23& 0.0581&  395&$+$\phn 72/$-$\phn 52\nl                 
A193     &  56&  56& 0.0490&  723&$+$\phn 78/$-$\phn 61&A2426    &  10&  10& 0.0886&  332&$+$\phn 80/$-$\phn 28\nl
A194     & 143&  39& 0.0184&  341&$+$\phn 57/$-$\phn 37&A2440    &  24&  24& 0.0913&  819&  upp. lim.  \nl                     
A229     &  34&  23& 0.1137&  506&    $+$165/$-$\phn 64&A2500    &  13&  13& 0.0904&  477&    $+$131/$-$\phn 54\nl                  
A256     &  15&  11& 0.0885&  545&    $+$107/$-$\phn 60&A2538a   &  23&  23& 0.0858&  326&$+$\phn 72/$-$\phn 59\nl             
A262     &  86&  82& 0.0169&  525&$+$\phn 47/$-$\phn 33&A2538b   &  18&  18& 0.0808&  410&$+$\phn 80/$-$\phn 48\nl                  
A295     &  47&  47& 0.0427&  359&$+$\phn 52/$-$\phn 32&A2554    &  28&  27& 0.1118&  840&    $+$131/$-$\phn 68\nl                  
A367a    &  13&  13& 0.0882&  394&    $+$150/$-$\phn 77&A2569    &  35&  35& 0.0816&  491&$+$\phn 86/$-$\phn 49\nl                  
A367b    &  13&  13& 0.0936&  479&    $+$207/$-$\phn 79&A2589    &  28&  28& 0.0423&  470&    $+$120/$-$\phn 84\nl                  
A399     &  87&  79& 0.0718& 1116&$+$\phn 89/$-$\phn 83&A2593    &  37&  37& 0.0424&  698&    $+$116/$-$\phn 69\nl                  
A400     &  98&  58& 0.0237&  599&$+$\phn 80/$-$\phn 65&A2634    & 217&  69& 0.0316&  700&$+$\phn 97/$-$\phn 61\nl                  
A401     & 107& 106& 0.0737& 1152&$+$\phn 86/$-$\phn 70&A2644    &  12&  12& 0.0693&  179&$+$\phn 48/$-$\phn 21\nl                  
A420     &  15&  15& 0.0860&  360&$+$\phn 70/$-$\phn 66&A2666    &  21&  21& 0.0280&  383&    $+$150/$-$\phn 75\nl                  
A426     & 200& 113& 0.0178& 1026&    $+$106/$-$\phn 64&A2670    & 235& 197& 0.0767&  852&$+$\phn 48/$-$\phn 35\nl                  
A458     &  32&  30& 0.1057&  736&$+$\phn 86/$-$\phn 58&A2715    &  13&  13& 0.1145&  463&    $+$152/$-$\phn 72\nl                  
A496     & 149&  55& 0.0325&  687&$+$\phn 89/$-$\phn 76&A2717    &  57&  55& 0.0498&  541&$+$\phn 65/$-$\phn 41\nl                  
A514     &  95&  81& 0.0714&  882&$+$\phn 84/$-$\phn 64&A2721    &  75&  74& 0.1152&  805&$+$\phn 74/$-$\phn 63\nl                  
A524     &  14&   8& 0.0797&  250&$+$\phn 62/$-$\phn 32&A2734    &  83&  80& 0.0625&  628&$+$\phn 61/$-$\phn 57\nl                  
A539     & 180& 160& 0.0284&  629&$+$\phn 70/$-$\phn 52&A2755    &  20&  20& 0.0957&  768&    $+$139/$-$\phn 84\nl              
A548NE   &  92&  62& 0.0397&  571&$+$\phn 54/$-$\phn 40&A2798    &  18&  18& 0.1130&  711&    $+$181/$-$101\nl                      
A548SW   &  74&  74& 0.0439&  583&$+$\phn 60/$-$\phn 37&A2799    &  36&  36& 0.0640&  422&$+$\phn 76/$-$\phn 57\nl                  
A569     &  39&  39& 0.0201&  327&$+$\phn 95/$-$\phn 39&A2800    &  31&  31& 0.0643&  404&$+$\phn 68/$-$\phn 74\nl                  
A576     &  48&  47& 0.0384&  945&$+$\phn 93/$-$\phn 88&A2819a   &  43&  33& 0.0876&  282&$+$\phn 50/$-$\phn 32\nl             
A634     &  15&  15& 0.0253&    0&  $+$223/$-$\phm{00}0&A2819b   &  49&  49& 0.0756&  410&$+$\phn 59/$-$\phn 44\nl                
A754     &  82&  77& 0.0535&  662&$+$\phn 77/$-$\phn 50&A2877    &  97&  93& 0.0248&887&$+$\phn 94/$-$\phn 68\nl                    
A978     &  57&  55& 0.0539&  535&$+$\phn 58/$-$\phn 39&A2911    &  30&  30& 0.0816&  547&    $+$159/$-$\phn 93\nl                  
A999     &  24&  24& 0.0317&  278&    $+$104/$-$\phn 49&A3093    &  22&  22& 0.0836&  440&$+$\phn 80/$-$\phn 56\nl                  
A1016    &  23&  23& 0.0318&  244&$+$\phn 43/$-$\phn 32&A3094    &  68&  67& 0.0677&  653&$+$\phn 77/$-$\phn 54\nl                  
A1060    & 144&  82& 0.0126&  610&$+$\phn 52/$-$\phn 43&A3111    &  12&  12& 0.0774&  159&$+$\phn 54/$-$\phn 28\nl                  
A1069    &  21&  21& 0.0662&  360&    $+$118/$-$\phn 59&A3122    &  91&  87& 0.0605&  775&$+$\phn 58/$-$\phn 51\nl              
A1142    &  43&  40& 0.0350&  486&$+$\phn 81/$-$\phn 41&A3126    &  41&  38& 0.0862& 1053&    $+$164/$-$108\nl                      
A1146    &  64&  57& 0.1422&  929&    $+$101/$-$\phn 83&A3128    & 186& 179& 0.0604&  789&$+$\phn 51/$-$\phn 44\nl                  
A1185    &  77&  21& 0.0300&  536&    $+$106/$-$\phn 56&A3142    &  21&  20& 0.1036&  737&$+$\phn 98/$-$\phn 63\nl                  
A1228    &  15&  15& 0.0354&  168&$+$\phn 47/$-$\phn 36&A3151    &  14&  14& 0.0662&  237&    $+$171/$-$\phn 40\nl
A1314    &  13&  13& 0.0329&  277&$+$\phn 83/$-$\phn 45&A3158    & 135& 123& 0.0597&  976&$+$\phn 70/$-$\phn 58\nl                  
A1631    &  71&  71& 0.0464&  702&$+$\phn 54/$-$\phn 46&A3194    &  31&  31& 0.0977&  805&$+$\phn 78/$-$\phn 53\nl                  
A1644    &  91&  84& 0.0467&  759&$+$\phn 61/$-$\phn 56&A3223    &  67&  66& 0.0603&  647&$+$\phn 67/$-$\phn 54\nl          
A1656    & 410& 283& 0.0233&  821&$+$\phn 49/$-$\phn 38&A3225    &  40&  36& 0.0563&  820& upp. lim.   \nl                          
A1775a   &  28&  10& 0.0759&  293&    $+$196/$-$133    &A3266    & 132& 128& 0.0599& 1107&$+$\phn 82/$-$\phn 65\nl                  
A1775b   &  25&  25& 0.0650&  478&    $+$117/$-$\phn 63&A3334    &  30&  30& 0.0970&  696&$+$\phn 91/$-$\phn 79\nl                  
A1795    &  85&  81& 0.0631&  834&$+$\phn 85/$-$\phn 76&A3354    &  57&  57& 0.0585&  358&$+$\phn 50/$-$\phn 45\nl                  
A1809    &  60&  59& 0.0789&  765&$+$\phn 79/$-$\phn 66&A3360    &  36&  34& 0.0849&  835&    $+$114/$-$\phn 82\nl                  
A1983    &  74&  74& 0.0452&  494&$+$\phn 43/$-$\phn 39&A3376    &  77&  75& 0.0465&  688&$+$\phn 68/$-$\phn 57\nl                  
A1991    &  23&  23& 0.0593&  631&    $+$147/$-$137    &A3381    &  29&  21& 0.0382&  293&    $+$110/$-$\phn 54\nl                  
A2029    &  91&  73& 0.0766& 1164&$+$\phn 98/$-$\phn 78&A3389    &  38&  38& 0.0272&  595&$+$\phn 63/$-$\phn 47\nl              
A2040    &  52&  28& 0.0454&  458&    $+$141/$-$102    &A3391    &  55&  18& 0.0553&  663&    $+$195/$-$112\nl                      
A2048    &  25&  25& 0.0972&  664&    $+$116/$-$\phn 65&A3395    & 105&  99& 0.0506&  852&$+$\phn 84/$-$\phn 53\nl                  
A2063    &  92&  92& 0.0350&  667&$+$\phn 55/$-$\phn 41&A3528    &  28&  28& 0.0536&  972&    $+$110/$-$\phn 82\nl                  
A2079    &  27&  26& 0.0662&  670&    $+$113/$-$\phn 67&A3532    &  43&  42& 0.0559&  738&    $+$112/$-$\phn 85\nl                  
A2092    &  17&  17& 0.0673&  536&    $+$129/$-$\phn 75&A3556    &  44&  43& 0.0476&  580&    $+$100/$-$\phn 73\nl                  
A2107    &  65&  65& 0.0415&  622&$+$\phn 71/$-$\phn 64&A3558    & 353& 341& 0.0480&  977&$+$\phn 39/$-$\phn 34\nl                  
A2124    &  65&  61& 0.0661&  878&$+$\phn 90/$-$\phn 72&A3559    &  37&  37& 0.0469&  456&$+$\phn 78/$-$\phn 44\nl                  
A2142    & 103&  86& 0.0907& 1132&    $+$110/$-$\phn 92&A3562    & 118& 100& 0.0478&  736&$+$\phn 49/$-$\phn 36\nl                  
A2151    & 101&  57& 0.0366&  751&$+$\phn 91/$-$\phn 69&A3571    &  70&  69& 0.0396& 1045&    $+$109/$-$\phn 90\nl
A2197    &  45&  45& 0.0308&  612&$+$\phn 56/$-$\phn 53&A3574    &  39&  35& 0.0158&  491&$+$\phn 73/$-$\phn 41\nl                  
A2199    &  70&  50& 0.0314&  801&$+$\phn 92/$-$\phn 61&A3651    &  78&  68& 0.0610&  626&$+$\phn 60/$-$\phn 53\nl
A2255    &  31&  25& 0.0824&\nodata &\nodata\nodata    &A3667    & 167& 154& 0.0566&  971&$+$\phn 62/$-$\phn 47\nl
A2256    &  87&  86& 0.0589& 1348&$+$\phn 86/$-$\phn 64&A3691    &  33&  29& 0.0881&  575&   upp. lim.         \nl        
A3693    &  15&  15& 0.0921&  478&    $+$107/$-$\phn 50&S463     &  84&  84& 0.0413&  608&$+$\phn 45/$-$\phn 41\nl
A3695    &  81&  74& 0.0903&  779&$+$\phn 67/$-$\phn 49&S753     &  32&  32& 0.0142&  536&    $+$127/$-$\phn 88\nl
A3703a   &  17&  17& 0.0743&  472&$+$\phn 87/$-$\phn 61&S805     & 119& 118& 0.0160&  541&$+$\phn 57/$-$\phn 43\nl
A3703b   &  13&  13& 0.0923&  554&    $+$122/$-$\phn 55&S987     &  33&  29& 0.0717&  677&    $+$141/$-$\phn 66\nl
A3705    &  40&  40& 0.0906&  877&$+$\phn 73/$-$\phn 74&AWM1     &  37&  14& 0.0287&  442&    $+$119/$-$\phn 41\nl
A3716N   &  43&  43& 0.0493&  466&$+$\phn 75/$-$\phn 58&AWM4     &  23&  23& 0.0328&  119&$+$\phn 89/$-$\phn 39\nl
A3716S   &  73&  69& 0.0458&  803&$+$\phn 58/$-$\phn 47&AWM7     &  33&  33& 0.0178&  864&    $+$110/$-$\phn 80\nl
A3733    &  41&  41& 0.0398&  608&    $+$109/$-$\phn 60&C67      &  27&  24& 0.0587&  581&    $+$136/$-$\phn 92\nl 
A3744    &  73&  57& 0.0390&  508&$+$\phn 74/$-$\phn 48&CL2335$-$\phn26&36&  29& 0.1249&  601&    $+$115/$-$\phn 63\nl                 
A3764    &  38&  35& 0.0766&  593&  upp. lim.          &DC0003$-$\phn50&35&  24& 0.0356&  348&$+$\phn 54/$-$\phn 40\nl                    
A3809    &  95&  71& 0.0631&  478&$+$\phn 62/$-$\phn 45&Eridanus &  54&  54& 0.0058&  264&$+$\phn 29/$-$\phn 27\nl                    
A3822    &  83&  77& 0.0769&  810&$+$\phn 89/$-$\phn 58&MKW1     &  15&  15& 0.0198&  227&$+$\phn 87/$-$\phn 37\nl                   
A3825    &  63&  59& 0.0760&  699&$+$\phn 79/$-$\phn 58&MKW3S    &  30&  30& 0.0450&  610&$+$\phn 69/$-$\phn 52\nl               
A3879    &  45&  40& 0.0679&  398&$+$\phn 60/$-$\phn 36&MKW4     &  51&  51& 0.0198&  525&$+$\phn 71/$-$\phn 48\nl                
A3921    &  31&  29& 0.0944&  490&    $+$126/$-$\phn 73&MKW5     &  20&  20& 0.0246&    0&    $+$135/$-$\phn\phn 0\nl                  
A4008    &  27&  27& 0.0558&  427&$+$\phn 64/$-$\phn 42&MKW6A    &  13&  13& 0.0257&  273&$+$\phn 83/$-$\phn 31\nl             
A4010    &  28&  28& 0.0966&  625&    $+$127/$-$\phn 95&MKW10    &  17&  17& 0.0201&  161&$+$\phn 42/$-$\phn 20\nl                  
A4053    &  16&  16& 0.0729&  615&    $+$212/$-$\phn 79&MKW12    &  16&  16& 0.0198&  233&    $+$101/$-$\phn 29\nl                  
A4067    &  29&  17& 0.0998&  499&    $+$123/$-$\phn 74&Ursa     &  57&  57& 0.0032&  128&$+$\phn 17/$-$\phn 11\nl                  
S84      &  19&  17& 0.1086&  329&$+$\phn 60/$-$\phn 25&Virgo    & 436& 179& 0.0038&  632&$+$\phn 41/$-$\phn 29\nl
S373     &  57&  57& 0.0499&  310&$+$\phn 30/$-$\phn 25&         &    &    &       &     &                     \nl  
\enddata
\end{deluxetable}

%% file: tab/tab3.tex
\begin{deluxetable}{lrrrrl}
\tablecolumns{6}
\tablewidth{0pc}
\tablecaption{The Results of Morphological Analysis}
\tablehead{
\colhead{Name} &
\colhead{$N_e$}    & 
\colhead{$N_l$}  & 
\colhead{$P_m$} &
\colhead{$P_F$}    & 
\colhead{References}\\
\colhead{(1)} &
\colhead{(2)} & 
\colhead{(3)} & 
\colhead{(4)} &
\colhead{(5)}    & 
\colhead{(6)}
}
\startdata
 A119           & 58 & 14 & 0.976 &   0.955& Fab105,788\nl
 A151b          &  6 &  5 & 0.370 &$>$0.999& Dre42,565\nl
 A194           & 38 & 25 & 0.644 &   0.989& Chi214,351\nl
 A229$^*$       & 21 &  8 & 0.395 &   0.997& E\nl
 A400           & 57 & 24 & 0.994 &   0.596& PCG;But57,665;Dre42,565\nl
 A496           & 54 & 17 & 0.991 &   0.396& Dre42,565\nl
 A524$^*$       &  5 &  5 & 0.977 &   0.905& E\nl
 A548NE$^*$     & 61 & 30 & 0.982 &   0.624& E;Dre95,284\nl 
 A1060 Hydra    & 81 & 54 & 0.993 &   0.395& Ric67,237;Ric77,237\nl
 A1775a         &  9 & 11 & 0.794 &   0.996& Oeg110,1\nl
 A2142$^*$      & 76 &  9 & 0.107 &   0.973& Oeg110,1\nl
 A2151 Hercules & 51 & 41 & 0.989 &   0.094& Dre95,284\nl
 A2634          & 64 & 22 & 0.427 &$>$0.999& Dr42,565\nl
 A2819a$^*$     & 25 & 10 & 0.840 &   0.995& E\nl
 A3381          & 16 &  8 & 0.581 &   0.957& Dre95,284\nl
 A3651$^*$      & 60 &  7 & 0.911 &$>$0.999& E\nl
 A3744$^*$      & 56 &  6 & 0.326 &   0.994& E\nl
 A3809$^*$      & 70 & 16 & 0.109 &   0.995& E\nl
 DC0003-50      & 22 & 11 & 0.864 &   0.974& Dre95,284\nl
 Virgo          &178 &244 & 0.128 &$>$0.999& Bin90,1681\nl
\tablecomments{
An asterisk indicates that spectral information is used. The
references for spectral features are the same as for galaxy
redshifts. We refer to Table 1 for the code used in the
references.
}
\enddata
\end{deluxetable}

%% file: tab/tab4.tex
\begin{deluxetable}{lcccrr}
\tablecolumns{6}
\tablewidth{0pc}
\tablecaption{The $\sigma$ Values for Multipeaked Clusters}
\tablehead{
\colhead{Name} &
\colhead{$N_p$}    & 
\colhead{$N_{mem}$}  & 
\colhead{z} &
\multicolumn{2}{c}{$\sigma$ (km/s)}\\
\colhead{(1)} &
\colhead{(2)} & 
\colhead{(3)} & 
\colhead{(4)} &
\multicolumn{2}{c}{(5)}
}
\startdata
A13      &  37&  37& 0.0950&  896& $+$\phn85/$-$73\nl
A87      &  19&  19& 0.0568&  511& $+$\phn98/$-$33\nl
A118     &  26&  26& 0.1155&  622& $+$\phn84/$-$57\nl
A168     &  75&  65& 0.0451&  435& $+$\phn31/$-$24\nl
A957     &  44&  44& 0.0436&  659& $+$\phn88/$-$56\nl
A1367    &  84&  84& 0.0216&  798& $+$\phn75/$-$68\nl
A1651    &  30&  30& 0.0844& 1006&    $+$118/$-$92\nl
A1736a   &  36&  34& 0.0347&  415& $+$\phn75/$-$42\nl
A1736bc  &  63&  63& 0.0461&  854& $+$\phn70/$-$52\nl
A2052    &  45&  45& 0.0356&  520& $+$\phn55/$-$47\nl
A2854    &  20&  20& 0.0622&  367& $+$\phn74/$-$76\nl
A3112    &  66&  62& 0.0757&  552& $+$\phn86/$-$63\nl
A3202    &  26&  26& 0.0698&  443& $+$\phn59/$-$37\nl
A3341    &  64&  64& 0.0379&  585& $+$\phn55/$-$38\nl
A3526    & 263& 156& 0.0114&  791& $+$\phn60/$-$62\nl
A3806    &  98&  81& 0.0777&  771& $+$\phn55/$-$50\nl
A4038    &  60&  60& 0.0302&  847& $+$\phn77/$-$49\nl
Pegasus  &  76&  76& 0.0140&  780& $+$\phn47/$-$52\nl
\enddata
\end{deluxetable}

%% file: tab/tab5.tex
\begin{deluxetable}{lcccrr}
\tablecolumns{6}
\tablewidth{0pc}
\tablecaption{The $\sigma$ Values for Multipeaked Clusters (Disjoined Peaks)}
\tablehead{
\colhead{Name} &
\colhead{$N_p$}    & 
\colhead{$N_{mem}$}  & 
\colhead{z} &
\multicolumn{2}{c}{$\sigma$ (km/s)}\\
\colhead{(1)} &
\colhead{(2)} & 
\colhead{(3)} & 
\colhead{(4)} &
\multicolumn{2}{c}{(5)}
}
\startdata
A13a            &  21&  21& 0.0972&  515&    $+$104/$-$81\nl
A13b            &  16&  16& 0.0919&  361& $+$\phn53/$-$35\nl
A87a            &  13&  13& 0.0558&  224& $+$\phn55/$-$24\nl
A118a           &  13&  13& 0.1137&  187& $+$\phn40/$-$28\nl
A118b           &  13&  13& 0.1172&  335&    $+$127/$-$64\nl
A168a           &  25&  25& 0.0469&  118& $+$\phn27/$-$16\nl
A168b           &  29&  29& 0.0436&  160& $+$\phn30/$-$19\nl
A168c           &  21&  21& 0.0454&   88& $+$\phn16/$-$13\nl
A957a           &  21&  21& 0.0416&  254& $+$\phn86/$-$41\nl
A957b           &  23&  23& 0.0457&  481&    $+$107/$-$66\nl
A1367a          &  68&  68& 0.0221&  570& $+$\phn72/$-$62\nl
A1651a          &  20&  20& 0.0863&  685&    $+$129/$-$99\nl
A1651c          &  10&  10& 0.0805&  363&    $+$182/$-$74\nl
A1736a          &  36&  34& 0.0347&  415& $+$\phn75/$-$42\nl
A1736b          &  37&  37& 0.0439&  569& $+$\phn54/$-$38\nl
A1736c          &  26&  26& 0.0493&  456& $+$\phn61/$-$40\nl
A2052a          &  28&  28& 0.0346&  207& $+$\phn41/$-$26\nl
A2854a          &  14&  14& 0.0619&  130& $+$\phn32/$-$26\nl
A3112b          &  13&  13& 0.0735&   86& $+$\phn33/$-$17\nl
A3202a          &  16&  16& 0.0708&  250& $+$\phn43/$-$26\nl
A3202b          &  10&  10& 0.0683&  179&    $+$133/$-$37\nl
A3341a          &  43&  43& 0.0390&  351& $+$\phn52/$-$39\nl
A3341b          &  21&  21& 0.0356&  209& $+$\phn30/$-$25\nl
A3526a          & 185& 121& 0.0108&  447& $+$\phn56/$-$36\nl
A3526b          &  78&  41& 0.0157&  289& $+$\phn59/$-$31\nl
A3806a          &  56&  56& 0.0762&  502& $+$\phn66/$-$56\nl
A3806b          &  29&  27& 0.0808&  297& $+$\phn47/$-$28\nl
A4038a          &  40&  40& 0.0285&  413& $+$\phn65/$-$40\nl
A4038b          &  20&  20& 0.0339&  337& $+$\phn72/$-$41\nl
Pegasus-a       &  37&  37& 0.0130&  288& $+$\phn43/$-$16\nl
Pegasus-b       &  27&  27& 0.0169&  223& $+$\phn51/$-$44\nl
\enddata
\end{deluxetable}

%% file: tab/tab6.tex
\begin{deluxetable}{rcc}
\tablecolumns{3}
\tablewidth{0pc}
\tablecaption{$\sigma$ vs. Abell Richness $R$ }
\tablehead{
\colhead{$R$} &
\colhead{$N$}    & 
\colhead{$\sigma$ (km/s)}\\ 
\colhead{(1)} &
\colhead{(2)} & 
\colhead{(3)}
}
\startdata
$-$1&6& 500$\pm61$\nl
0&22& 485$\pm45$\nl
1&69& 587$\pm28$\nl
2&34& 770$\pm45$\nl
3&7& 807$\pm64$\nl
\enddata
\end{deluxetable}